\documentclass[12pt,nofootinbib]{revtex4-1}

\usepackage[utf8]{inputenc}
\usepackage{amsmath}
\usepackage{amssymb}
\usepackage{amsfonts}
\usepackage{graphicx}
\usepackage{color}
\usepackage{amsbsy}
\usepackage{url}
\usepackage{enumerate}
\usepackage[vcentermath,enableskew]{youngtab}
\usepackage{verbatim}
\usepackage[normalem]{ulem}
\usepackage{cancel}
\usepackage{bm,bbm}
\usepackage{slashed}

\makeatletter
\@addtoreset{equation}{section}
\renewcommand{\theequation}{\thesection.\arabic{equation}}
\makeatother

\newcommand{\refer}[1]{(\ref{#1})}
\newcommand{\diff}{\mathrm{d}}
\newcommand{\comm}[2]{\bigl[ #1, #2 \bigr]}

\newcommand{\lineint}{\int\limits_{-\infty}^{\infty}\diff}

\newcommand{\Tr}{\mathrm{Tr}}

\newcommand{\sech}{\mbox{sech}}

\def\be{\begin{eqnarray}}
\def\ee{\end{eqnarray}}
\def\p{\partial}

\begin{document}

\begin{flushright}
YGHP-17-05
\end{flushright}

\title{
Non-Abelian Gauge Field Localization on Walls\\
and
Geometric Higgs Mechanism
}

\author{Masato Arai$^1$, Filip Blaschke$^2$, Minoru Eto$^3$ and Norisuke Sakai$^4$\ \\\ }
\affiliation{
$^1$Faculty of Science, Yamagata University, 
Kojirakawa-machi 1-4-12, Yamagata,
Yamagata 990-8560, Japan\\
$^2$Faculty of Philosophy and Science, Silesian University in Opava, Bezru\v{c}ovo n\'am. 1150/13, 746~01 Opava, Czech Republic\\
$^3$Department of Physics, Yamagata University, 
Kojirakawa-machi 1-4-12, Yamagata,
Yamagata 990-8560, Japan\\
$^4$Department of Physics, and Research and 
Education Center for Natural Sciences, 
Keio University, 4-1-1 Hiyoshi, Yokohama, Kanagawa 223-8521, Japan
}
\email{arai(at)sci.kj.yamagata-u.ac.jp\\
fblasch(at)post.cz\\
meto(at)sci.kj.yamagata-u.ac.jp\\
norisuke.sakai(at)gmail.com}

\begin{abstract}
Combining the semi-classical localization mechanism for gauge fields with $N$ domain wall background in a simple $SU(N)$ gauge theory in five space-time dimensions we investigate the geometric Higgs mechanism, where a spontaneous breakdown of the gauge symmetry comes from splitting of domain walls. The mass spectra are investigated in detail for the phenomenologically interesting case $SU(5) \to SU(3)\times SU(2)\times U(1)$ which is realized on a split configuration of coincident triplet and doublet of domain walls. We derive a low energy effective theory in a generic background using the moduli approximation, where all non-linear interactions between effective fields are captured up to two derivatives. We observe novel similarities between  domain walls in our model and D-branes in superstring theories.
\end{abstract}

\maketitle

\newpage

%


\section{Introduction}

One of the most puzzling features of the Standard Model (SM) 
is the lack of explanation of  the gauge hierarchy problem. 
To solve this problem, apart from other popular ideas such as supersymmetry 
\cite{Witten:1981nf, Sakai:1981gr, Dimopoulos:1981zb, Dimopoulos:1981yj}, 
and composite (Technicolor) models 
\cite{Weinberg:1979bn, Susskind:1978ms}, the brane world 
scenario has been invoked in various forms \cite{ADD, RS, RS2, Arkani-Hamed,Antoniadis:1998ig}.

The possibility of dynamical realization of the brane world idea via a domain wall was recognized 
quite early \cite{Rubakov}.
A long-lasting obstacle for serious investigations of brane-world 
scenarios by domain walls, however, was the localization of gauge fields. 
Naive attempts to localize gauge fields on the domain wall 
with the Higgs phase in the bulk give no massless 
gauge fields in the effective 
theory \cite{Dvali:1996xe, ADD, Maru:2003mx} (see also \cite{Germani:2011cv, Dvali:2000rx, Dubovsky:2001pe, Akhmedov:2001ny} for related studies).
The so-called Dvali-Shifman (DS) mechanism \cite{Dvali:1996xe} 
is a popular way to get around the problem, inspired by a 
non-perturbative feature of the non-Abelian gauge theories -- 
the confinement. 
However,  it has not been proven whether non-Abelian
gauge theories which exhibit the confinement in the bulk exist in $(d+1)$-dimensional spacetime with $d \ge 4$.

It was pointed out in Ref.~\cite{Ohta} that one can implement 
the gauge field localization more easily in a semi-classical way.
If the gauge coupling depends on the 
extra-dimensional coordinate in such a way that it rapidly 
diverges away from the brane (semi-classical picture of 
confinement) while remaining finite in the vicinity of the brane, 
it effectively provides a confining vacuum for zero modes of 
gauge fields with the four-dimensional gauge invariance intact. 
The mechanism is realized by a 
field-dependent gauge kinetic term \cite{Ohta}.
This arises naturally 
in ${\mathcal N}=2$ supersymmetric gauge theories in five spacetime 
dimensions in the form of the so-called prepotential 
\cite{Seiberg:1996bd, Morrison:1996xf}. 
In this framework, we have 
constructed models of non-Abelian gauge fields localized 
around domain walls and worked out nonlinear interactions of 
moduli fields \cite{Us1, Us2}.

In this paper, we investigate the Higgs mechanism caused by the domain walls.
In the previous works \cite{Us1, Us2} our studies were focused on how to localize the massless
non-Abelian gauge fields on the walls. In contrast, in the present paper, we aim at figuring out how the massless
gauge fields get non-zero masses in the framework  \cite{Ohta,Us1,Us2}.
Either by DS or Ohta-Sakai (OS) mechanism, the localization of non-Abelian gauge fields occurs due to the confining phase in the bulk. This has
many similarities with the localization of gauge fields on D-branes in superstring theories.
Indeed, we found in our previous works \cite{Us1,Us2} that $N$ 
coincident domain walls are needed to have massless $SU(N)$ 
gauge fields inside the domain walls.
Therefore, we naturally expect that the Higgs mechanism also 
goes similarly to low energy effective theory on D-branes, 
and we will show it is indeed so.

It is often the case that a non-Abelian global symmetry 
is realized in the coincident wall configuration. 
It has been found previously, that splitting of domain 
walls can break the global symmetry and the moduli fields 
corresponding to the wall positions become massless 
Nambu-Goldstone (NG) bosons associated to the symmetry 
breaking \cite{Eto:2008dm, Eto:2009zv}. 
When non-Abelian gauge fields couple to the global symmetry, one naively expects that
they will absorb these moduli fields and become massive. 
If this is the case, a splitting of positions of domain walls in the 
five-dimensional theory can induce a spontaneous breakdown 
of non-Abelian gauge symmetry in the effective theory on 
domain walls. 
In other words, the moduli fields corresponding to the wall 
positions play the role of the Higgs field in the effective 
field theory. 
Since the geometrical data such as wall positions provide 
scalar fields realizing Higgs phenomenon, we call this 
mechanism as the \emph{geometric Higgs mechanism}.

In our previous works, we have observed geometric Higgs mechanism 
\cite{Us1, Us2} indirectly through effective Lagrangian.
The purpose of this paper is to give direct study of the geometric 
Higgs mechanism from the 5-dimensional point of view in detail. 
Since the would-be NGs are not homogeneously distributed as 
they are affected by the domain wall background, the geometric 
Higgs mechanism is not as straightforward as the standard Higgs 
mechanism in homogeneous Higgs vacuum.
We will study physical spectrum via mode equations for all fields in 
detail 
and show that the gauge fields associated with the broken gauge symmetry absorb the localized NGs and get non-zero masses.

Furthermore, we calculate the four-dimensional low-energy 
effective Lagrangian in the arbitrary domain wall background 
in the so-called moduli approximation \cite{Manton}. 
This effective Lagrangian captures full non-linear interactions 
between moduli fields up to two derivative terms, which we write 
down in a closed form. With the effective Lagrangian, we give a proof of
the geometric Higgs mechanism from the perspective of low energy
effective theory on the domain walls.

Lastly, many similarities between domain walls and D-branes 
have been shown in the literature. 
For example, D1-D3-like configuration was found in 
\cite{Caroll, Bowick:2003au, Witten:1997ep, Kogan:1997dt, 
Campos:1998db, Shifman:2004dr, Sakai2, Tong2, Arai:2016sdz, Arai:2016kur}. 
Furthermore, the low-energy effective theory on 
domain walls was found to be similar to that on D-branes 
\cite{Gauntlett:2000de, Shifman:2002jm, Shifman:2003uh, Hanany:2003hp}. 
In our work, we find new evidence for the correspondence 
between domain walls and D-branes. 
Like in D-branes, the number of coincident domain walls 
determines the rank of the gauge group. 
In addition, the masses of gauge bosons  
are proportional to the distance between walls, at least when 
they are close. 
As a result, our model further 
strengthens the notion of domain-wall-D-brane 
correspondence.

The paper is organized as follows. In Sec.~\ref{sec:3} we 
present an $SU(N)$ gauge theory with two adjoint scalar fields. 
In Sec.~\ref{sec:walls} we construct $N$ domain walls 
and discuss the ungauged fluctuation spectrum.
In Sec.~\ref{sec:threetwo} we turn on the gauge interactions 
and analyze the spectrum of fluctuations around the 3-2 split 
background in $N=5$ model to demonstrate the geometric Higgs mechanism. 
Sec.~\ref{sc:effective_action2} is devoted to the low-energy 
effective Lagrangian in four dimensions with the moduli 
approximation. 
Lastly, sec.~\ref{sec:discussion} is devoted to summary and 
future prospects. 
In Appendix~\ref{app:KKAH} we present the Kaluza-Klein spectrum of Abelian-Higgs model on $\mathbb{R}^{3,1}\times S^1/Z_2$, which we consider as a toy model to our theory, while we have collected several identities 
useful to compute effective Lagrangian 
in the Appendix~\ref{app:A}.

\section{The model}\label{sec:3}

Let us consider a (4+1)-dimensional $SU(N)$ gauge theory with 
two adjoint scalars $\hat T$ and $\hat S$ transforming as 
$\hat T \to U_{N} \hat T U_{N}^{\dagger}$ and $\hat S \to U_{N} \hat S U_{N}^{\dagger}$ with $U_{N} \in SU(N)$, 
and two singlets $T^0$ and $S^0$. 
We combine both adjoints and singlets into 
Hermitian $N\times N$ matrices 
$T \equiv \hat T +\mathbf{1}_N\frac{T^0}{N}$ and 
$S \equiv \hat S +\mathbf{1}_N\frac{S^0}{N}$. 
The Lagrangian is given as
\begin{equation}\label{eq:lagrangian}
{\mathcal L} = {\mathcal L}_{\rm B}+{\mathcal L}_{\rm OS}\,.
\end{equation} 
The first part ${\mathcal L}_{\rm B}$ contains kinetic and 
potential terms for bosons except for the $SU(N)$ gauge kinetic term as
\begin{align}
{\mathcal L}_{\rm B} &= \Tr\bigl[D_M T D^M T
+D_M S D^M S \bigr] -V\,, \\ \label{eq:volpot}
V &= \Tr\Bigl[\lambda^2\bigl(v^2 \mathbf{1}_N-T^2
-S^2\bigr)^2+\Omega^2S^2 - \xi \comm{T}{S}^2\Bigr]\,,
\end{align}
where
$\lambda$ and $\xi$ are coupling constants and where $\Omega$ is a 
mass parameter for $S$. We use mostly negative metric signature. The covariant derivatives are 
defined by
\begin{equation}
D_M T = \partial_M T + i\comm{A_M}{T}\,,\qquad
D_M S = \partial_M S + i\comm{A_M}{S}\,. 
\label{eq:cov-D}
\end{equation} 
The potential \refer{eq:volpot} is chosen not for its generality, but rather to ensure analytic solutions for both the background solution and most of the fluctuation spectra. This will help in subsequent sections to keep the discussion as simple as possible, without sacrificing the generality of our results as more generic potentials than \refer{eq:volpot} would make no qualitative difference.

The field-dependent gauge kinetic term ${\cal L}_{\rm OS}$ 
is given in the form
\begin{equation}\label{eq:ohta}
{\mathcal L}_{\rm OS}  = -  \Tr\bigl[F(S) G_{MN}G^{MN}\bigr]\,,
\end{equation} 
where  $F(S)$ is an arbitrary polynomial function of $S$, and 
$G_{MN} = 
\partial_MA_N-\partial_N A_M +i \comm{A_M}{A_N}$ is the field 
strength of $SU(N)$ gauge fields. 
The gauge transformation is defined by
$A_{\mu} \to U_N  A_{\mu}U_N^{\dagger} +i \partial_\mu U_N U_N^{\dagger}$.
The field-dependent gauge coupling term ${\mathcal L}_{\rm OS}$ 
is responsible for localization of gauge fields on the world-volume 
of domain walls in the background $T$ and $S$ fields. 
In the original work \cite{Ohta}, the function $F$ is restricted 
to be a linear function by the supersymmetry,
but in this work we do not impose supersymmetry and, for convenience, 
we take
\be
F(S) = a S^2,
\label{eq:ext_OS_factor}
\ee 
where we assume $a$ to be real and positive.
All the arguments below are not qualitatively changed if we 
consider the linear function $F=aS$ as in the original \cite{Ohta}. 
The reason why we take a quadratic function is to ensure positiveness 
of the gauge kinetic term.
The mass dimensions of the fields and parameters are summarized 
in Table \ref{tab:1}.
\begin{table}
\begin{center}
\begin{tabular}{c|ccc|cccccc}
fields and parameters & $A_M$ & $T$ & $S$  & $\Omega$ & $\lambda$ & $v$ & $\bar v$ & $\xi$ & $a$ \\
\hline
mass dimension & $1$ & $\frac{3}{2}$ & $\frac{3}{2}$ & $1$ & $-\frac{1}{2}$ & $\frac{3}{2}$ & $\frac{3}{2}$ & $-1$ & $-2$ 
\end{tabular}
\caption{The mass dimensions of the fields and parameters.}
\label{tab:1}
\end{center}
\end{table}

The equations of motion for the above model are 
\begin{align}
&D_M\left\{G^{MN},S^2\right\} - \frac{1}{N}\Tr\left[D_M\left\{G^{MN},S^2\right\}\right] \mathbf{1}_N 
= \frac{i}{a} \left(\left[D^NT,T\right] + \left[D^NS,S\right]\right),
\label{eq:fullEOM1}\\
&D^2T = \lambda^2\left\{T, v^2\mathbf{1}_N - T^2 - S^2\right\} + \xi \left[S,\left[T,S\right]\right],
\label{eq:fullEOM2}\\
&D^2S = \lambda^2\left\{S, v^2\mathbf{1}_N - T^2 - S^2\right\} + \xi \left[T,\left[S,T\right]\right] - \Omega^2S 
+ \frac{a}{2}\left\{S,G^2\right\},
\label{eq:fullEOM3}
\end{align}
with $D^2 = D_MD^M$ and $G^2 = G_{MN}G^{MN}$.

The potential $V$ in Eq.~\refer{eq:volpot} has 
a number of discrete vacua,  
\begin{gather}
T =  v \Lambda\,, \hspace{5mm} S = \mathbf{0}_N\,,
\label{eq:vacua}
\end{gather}
where $\Lambda^2 = \mathbf{1}_{N}$. Without loss of generality, 
by using the $SU(N)$ symmetry, 
we can diagonalize it as 
\begin{eqnarray}
\left<k,N-k\right>\text{\ vacuum}:\ \Lambda = {\rm diag}
(\,\underbrace{1,1, \cdots,1}_k,\, \underbrace{-1,-1,\cdots,- 1}_{N-k}\,).
\end{eqnarray}
The two vacua $\left<N,0\right>$ and $\left<0,N\right>$ are 
$SU(N)$ preserving vacua, which we will use as boundary conditions 
to obtain background domain wall solutions. All the remaining 
vacua partially break $SU(N)$. The breaking pattern of $SU(N)$ 
depends on $k$ as 
\begin{eqnarray}
SU(N) \to S\left[U(N-k) \times U(k)\right],\quad (k=0,1\cdots,N).
\end{eqnarray}

In order to find mass spectrum of each vacuum, let us first 
replace $F$ by 
\begin{eqnarray}
F_\varepsilon \equiv aS^2 + \frac{1}{4g_\varepsilon^2},
\end{eqnarray}
where $g_\varepsilon$ is a fictitious $SU(N)$ gauge coupling.
We reproduce the original gauge kinetic term at the limit 
$g_\varepsilon \to \infty$.
Since $S = 0$ at the vacua, we have an ordinary gauge kinetic 
term with $F_\varepsilon = \frac{1}{4g_\varepsilon^2}$.
In the $SU(N)$ preserving vacua, the masses of $T$ and $S$ are 
$m_T = \sqrt{2}\, \lambda v$ and $m_S = \Omega$.
The $SU(N)$ gauge fields are unbroken, hence they are massless.
The mass spectrum in the $\left<k,N-k\right>$ vacuum is the 
following. 
Similarly to the unbroken vacua, the $k$ by $k$ and $N-k$ by 
$N-k$ block diagonal elements of $T$ and $S$ are massive 
with masses $\sqrt{2}\, \lambda v$ and $\Omega$.
The remaining elements in off-diagonal blocks are nothing but 
the Nambu-Goldstone (NG) zero modes. 
The corresponding off-diagonal elements of the gauge fields 
absorb these NG bosons by the standard Higgs mechanism to have 
mass $\sqrt2 g_\varepsilon v$, whereas 
the gauge fields for the unbroken part $S[SU(k)\times SU(N-k)]$ 
remain massless.

Now, let us send $g_\varepsilon \to \infty$ and go back to the original model. The masses of $T$ and $S$, and also
the unbroken gauge fields are not affected by $g_\varepsilon$. Therefore, the block-diagonal components of $T$ and $S$
maintain their masses $\sqrt{2}\, \lambda v$ and $\Omega$ while those of the gauge fields remain massless.
On the other hand, the off-diagonal massive gauge bosons get frozen as their masses become infinitely large $g_\varepsilon v \to \infty$. 
At the same time, the unbroken $S[U(k)\times U(N-k)]$ gauge interaction has the infinitely large coupling constant $g_\varepsilon \to \infty$.
We interpret this as a semi-classical manifestation 
of confining vacua.
As we will see below, we can manifestly show that, thanks to the infinite gauge coupling, not only the massless gauge fields \cite{Ohta} 
but also the massive vector bosons localize on/between domain walls. 

In short, we insist that 
there are no light scalar fields in any vacua. They are heavy since their masses are of the five-dimensional mass scale $M_5$ which we assume
very large compared to four-dimensional mass scales.
Furthermore, the gauge fields are either confined or infinitely heavy. Therefore, no light degrees of freedom
exist in any vacua from five-dimensional viewpoint. This property should be important for the purpose of constructing
phenomenological models, though it is out of the scope of this paper.

\section{Multiple domain walls}\label{sec:walls}

\subsection{Background domain wall solutions}

Let us look for static $y\equiv x^4$-dependent domain wall solutions to Eqs.(\ref{eq:fullEOM1}) -- (\ref{eq:fullEOM3}).
 Setting $\partial_\mu = 0$ ($\mu=0,1,2,3$) and $A_M = 0$, the equations of motion reduce to
\be
T'' &=& -\lambda^2\left\{T, v^2\mathbf{1}_5 - T^2 - S^2\right\} - \xi \left[S,\left[T,S\right]\right],
\label{eq:feom1}\\
S'' &=& -\lambda^2\left\{S, v^2\mathbf{1}_5 - T^2 - S^2\right\} - \xi \left[T,\left[S,T\right]\right] + \Omega^2S.
\label{eq:feom2}
\ee
We solve these with the boundary conditions 
\be
T \to \pm v {\bf 1}_N,\quad S \to 0,\qquad \text{as}
\quad y \to \pm \infty.
\label{eq:bound_cond}
\ee
Note that these equations correspond to a non-Abelian extension 
of the well-known two-scalar MSTB model (named after Montonen, 
Sarker, Trullinger and Bishop), solutions of which 
have been studied in detail \cite{Montonen, Sarkar, Ito}.
Denoting the solution of MSTB model as  
the 1 by 1 scalar fields ${\cal T}(y)$ and ${\cal S}(y)$, we can 
immediately get domain wall solutions of our model by embedding 
${\cal T}(y)$ and ${\cal S}(y)$ into the matrices $T$ and $S$.

In the MSTB model, two types of domain wall-like solutions are known. The first type is
\be
{\cal T} = v \tanh v\lambda (y-y_0),\qquad 
{\cal S} = 0,
\label{eq:back_false}
\ee
which is known to be stable only in the parameter region $\Omega \ge v \lambda$.
However, we want $S$ field to condense inside the domain 
wall for trapping zero modes of gauge fields by Ohta-Sakai mechanism \cite{Ohta}. 
Therefore, this solution is not suitable for our purposes.

The second type, which is supported in the parameter region $\Omega < v\lambda$, has two different solutions, namely
\be
{\cal T} = v \tanh \Omega (y-y_0),\qquad
{\cal S} = \pm\, \bar v\, \sech\, \Omega (y-y_0),
\label{eq:back_true}
\ee
where we have defined
\begin{equation}
\bar v^2 \equiv v^2 -\frac{\Omega^2}{\lambda^2} > 0\,.
\end{equation}
The width of the wall is of order $\Omega^{-1}$.
One can choose either $+$ or $-$ discrete moduli, but we will use $+$ solution in what follows
for concreteness\footnote{
This is the main reason for choosing the quadratic 
function in Eq.~(\ref{eq:ext_OS_factor}). If $F(S)$ is linear 
as is the original work \cite{Ohta},
the solution with minus sign implies wrong sign of the kinetic term, and leads to instability of gauge interaction.
}.
The general domain wall solution with the $SU(N)$ unbroken vacua at $y \to \pm \infty$
can be constructed by embedding these into $T$ and $S$ as
\begin{equation}\label{eq:gensol3}
T  = v\tanh\Omega\bigl( y \mathbf{1}_N-Y\bigr)\,, \hspace{5mm}
S  = \bar v\,\sech\, \Omega \bigl( y\mathbf{1}_N-Y\bigr)\,,
\end{equation}
where a single $N\times N$ Hermitian matrix $Y$ contains all the free parameters of the solution.
Stability of this solution can be shown as follows.
Firstly, we can construct the Bogomol'nyi completion of energy density as 
\begin{align}
{\mathcal E} & = \Tr\Bigl[\Bigl\{\partial_y T
-\frac{\Omega}{2v^3}\bigl(2v^2+\bar\Omega^2\bigr)
\bigl(v^2\mathbf{1}_N-T^2\bigr)+\frac{\Omega}{2v}
S^2\Bigr\}^2+\Bigl(\partial_yS +\frac{\Omega}{2v}\bigl(
TS+ST\bigr)\Bigr)^2 \nonumber \\ \label{eq:bogone}
& +\frac{\bar v^4}{4v^2}(4v^2\lambda^2-\Omega^2)
\Bigl(\frac{T^2}{v^2}+\frac{S^2}{\bar v^2}
-\mathbf{1}_N\Bigr)^2
-\xi\comm{T}{S}^2\Bigr] + {\mathcal E}_0\,,
\end{align}
with the bound
\begin{equation}
{\mathcal E} \geq {\mathcal E}_0 \equiv \Tr\left[\frac{\Omega}{v^3}
\bigl(2v^2+\bar v^2\bigr)\partial_y\Bigl(v^2T
-\frac{1}{3}T^3\Bigr)-\frac{\Omega}{v}\partial_y\bigl(
TS^2\bigr)\right]\,.
\end{equation}
The bound is saturated when the energy equals the tension of 
the domain walls
\begin{equation}
E = \int_{-\infty}^{\infty}\diff y\, {\mathcal E}_0 
= N \times T_W\,,\quad T_W \equiv \frac{4\Omega}{3}\bigl(2v^2+\bar v^2\bigr)\,,
\label{eq:wall_tension}
\end{equation}
which implies the BPS equations
\begin{gather}
\partial_y T =\frac{v \Omega}{\bar v^2}S^2\,, 
\hspace{5mm} \partial_y S = -\frac{\Omega}{2v}\bigl(TS+ST\bigr)\,, 
\hspace{5mm} \frac{T^2}{v^2}+\frac{S^2}{\bar v^2} 
= \mathbf{1}_N\,, \hspace{5mm} \comm{T}{S} = 0\,.
\end{gather}
One can easily show that  $T$ and $S$ given in Eq.~(\ref{eq:gensol3}) solve these BPS equations.

\subsection{Domain walls in the global $SU(N$) model}
\label{sec:dwzm}

Let us figure out  physical meaning of the parameters 
contained in the $N\times N$ Hermitian matrix $Y$. 
For that purpose, only  in this subsection, we will consider the global $SU(N)$ model by turning off the gauge interaction:
\begin{align}
\tilde {\mathcal L}_{\rm B} &= \Tr\bigl[\partial_M T \partial^M T
+\partial_M S \partial^M S \bigr] -V\,, \label{eq:L_scalar_model}\\ 
V &= \Tr\Bigl[\lambda^2\bigl(v^2 \mathbf{1}_N-T^2
-S^2\bigr)^2+\Omega^2S^2 - \xi \comm{T}{S}^2\Bigr]\,.
\label{eq:P_scalar_model}
\end{align}
The $SU(N)$ symmetry is now a  global symmetry, and $T$ and $S$ in Eq.~(\ref{eq:gensol3})
are still solutions.
In the global $SU(N)$ model the parameters in $Y$ are all physical zero modes.
Since $Y$ is Hermitian, one can always diagonalize it  by an $SU(N)$ transformation as
$Y = {\rm diag}(y_1,y_2,\cdots,y_{N-1},y_N)$. We can set $y_1 \le y_2 \le \cdots \le y_N$ 
without loss of generality.
The solution is then of the form
\begin{align}\label{eq:gensol1}
T & = v\, \mbox{diag}\bigl(
\tanh\Omega(y-y_{1}), \ldots ,\tanh\Omega(y-y_{N})\bigr)\,, \\ \label{eq:gensol2}
S & =\bar v\, \mbox{diag}\bigl(
\sech\,\Omega(y-y_{1}), \ldots ,\sech\,\Omega(y-y_{N})\bigr) \,.
\end{align}
Now, it is manifest that the eigenvalues $\{y_i\}$ correspond to positions of the domain walls in the $y$ direction.
So, we have $N$ domain walls.

Let us next consider small fluctuation for $Y = Y_0 + \delta Y$ around a given $Y = Y_0$.
These fluctuations are zero modes because the shift does not change the energy of the solution.
When all the eigenvalues of $Y_0$ are different, the global $SU(N)$ symmetry is broken to the maximal Abelian subgroup $U(1)^{N-1}$.
Therefore $(N^2-1) - (N-1) $ out of $N^2$ zero modes in $Y$ are Nambu-Goldstone (NG) modes for $SU(N) \to U(1)^{N-1}$. 
We also have one NG mode for the broken translational symmetry and
$N-1$ quasi Nambu-Goldstone (qNG) modes associated with
the relative distance of $N$ domain walls. 
In the opposite case where all the eigenvalues of $Y_0$ are the same, the $N$ walls are all coincident and $SU(N)$ symmetry 
remains intact. There is only one NG for the broken translational symmetry which corresponds to ${\rm Tr}[\delta Y]$ and the remaining 
$N^2-1$ are qNG. The similar counting can be done for other cases.

\begin{figure}
\begin{center}
\includegraphics[width=16cm]{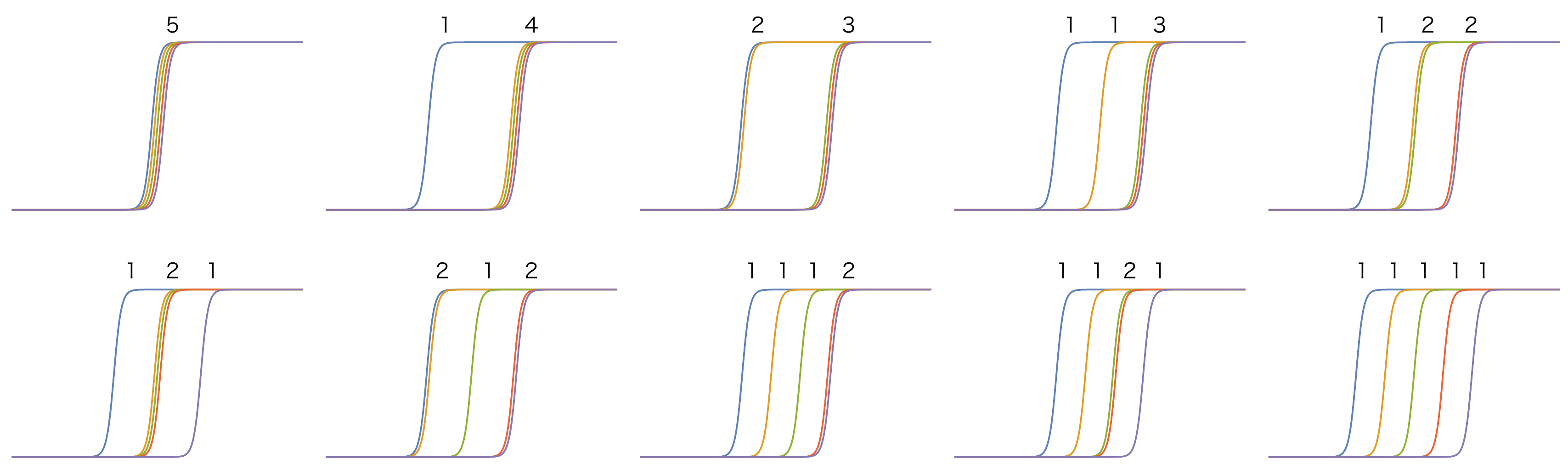}
\caption{Amplitudes of the diagonal component of $T$ for ten different patters of five domain walls in $SU(5)$ model.}
\label{fig:su5_10walls}
\end{center}
\end{figure}

To be concrete, let us consider $N=5$ in the rest of this subsection. 
Depending on the choice of values of the eigenvalues of $Y$, we have 
10 different patterns of domain walls as shown in Fig.~\ref{fig:su5_10walls}. From among those configurations, we concentrate on
$Y=Y_0$ with $y_1 = y_2 = y_3 \equiv {\cal Y}_3 <  
{\cal Y}_2 \equiv y_4 = y_5$. The domain walls connects the three vacua $\left<0,5\right>$, $\left<3,2\right>$, and $\left<5,0\right>$ ordered 
from left to right.
The $SU(5)$ symmetry is intact at both two vacua $\left<0,5\right>$ and $\left<5,0\right>$ but it breaks down to $SU(3) \times SU(2) \times U(1)$
in the middle $\left<3,2\right>$ vacuum. 
The number of NG modes for this partial symmetry breaking is
$24 - (8+3+1) = 12$.
This can easily be seen as follows. First, we divide  the background configuration into two parts:  $SU(5)$ unbroken part
and $SU(5)$ broken part as
\be
\left(
\begin{array}{c}
T\\
S
\end{array}
\right) 
&=&
\left(
\begin{array}{c}
\left(
\begin{smallmatrix}
\tau_3{\bf 1}_3 & 0\\
0 & \tau_2{\bf 1}_2
\end{smallmatrix}
\right)
\\
\left(
\begin{smallmatrix}
\sigma_3{\bf 1}_3 &0\\
0& \sigma_2{\bf 1}_2
\end{smallmatrix}
\right)
\end{array}
\right) \nonumber\\
&=& \label{eq:division}
\mathbf{1}_5 \otimes
\left(
\begin{array}{c}
\frac{3\tau_3+2\tau_2}{5}\\
\frac{3\sigma_3+2\sigma_2}{5}
\end{array}
\right)
+ \left(
\begin{array}{cc}
\frac{2}{5}\mathbf{1}_3 & \\
& -\frac{3}{5}\mathbf{1}_2
\end{array}
\right)
\otimes
\left(
\begin{array}{c}
\tau_3 - \tau_2\\
\sigma_3 - \sigma_2
\end{array}
\right)\,,
\ee
where we define
\be
\tau_i \equiv v \tanh \Omega(y-{\cal Y}_i),
\quad \sigma_i \equiv \bar v\,\sech\,\Omega(y-{\cal Y}_i),\quad (i=2,3)\,.
\label{eq:wall22}
\ee
Moreover, 
an infinitesimal global $SU(5)$ transformation can be parametrized as
\be
U_5 = \mathbf{1}_5 + i \left(
\begin{array}{cc}
\alpha_3 & \tilde \alpha \\
\tilde \alpha^\dagger & \alpha_2
\end{array}
\right)\,,
\label{eq:infinitesimal_SU5)}
\ee
where $\alpha_3$ and $\alpha_2$ are 
$3\times3$ and $2\times2$ infinitesimal Hermitian matrices with ${\rm Tr}\alpha_3 + {\rm Tr}\alpha_2 = 0$ belonging to $SU(3)\times SU(2) \times U(1)$, while
$\tilde \alpha$ is a 3 by 2 complex matrix containing the 12 broken generators. Applying it to Eq.~\refer{eq:division} we obtain
\be
\delta
\left(
\begin{array}{c}
T\\
S
\end{array}
\right) 
= \left(
\begin{array}{c}
\left(
\begin{smallmatrix}
0 & i\tilde\alpha(\tau_3-\tau_2)\\
-i\tilde\alpha^\dagger(\tau_3-\tau_2) & 0
\end{smallmatrix}
\right)\\
\left(
\begin{smallmatrix}
0 & i\tilde\alpha(\sigma_3-\sigma_2)\\
-i\tilde\alpha^\dagger(\sigma_3-\sigma_2) & 0
\end{smallmatrix}
\right)
\end{array}
\right)
=
\left(
\begin{array}{cc}
0 & i\tilde\alpha\\
-i\tilde\alpha^\dagger & 0
\end{array}
\right)
\otimes
\left(
\begin{array}{c}
\tau_3 - \tau_2\\
\sigma_3 - \sigma_2
\end{array}
\right)\,.
\label{eq:SU(5)_global_transf}
\ee
Thus the 12 zero modes in $\tilde \alpha$ are nothing but the NG modes.
We also have $3^2 + 2^2 -1 =12$ qNG modes living in the 3 by 3 top-left and 2 by 2 bottom-right corner of $\delta Y$. 
Adding the translational zero mode, we again have $12 + 12 + 1 = 25$ zero modes in total.
It is important to observe that physics such as massive 
spectra and the character of massless modes (NG boson or qNG boson) 
differ depending on different values of moduli parameters. 
However, the total number 
of massless modes (NG and qNG together) remains the same 
irrespective of the value of moduli parameters \cite{Eto:2008dm}.

Let us verify mass spectra and wave functions of each mode by 
considering small fluctuations around a background configuration. 
We again take the 3-2 splitting background solution  (it is a 
straightforward task to generalize the following to other cases)
\be
T &=& \left(
\begin{array}{cc}
\tau_3(y) \mathbf{1}_3 & 0\\
0 & \tau_2(y) \mathbf{1}_2
\end{array}
\right)
+ \left(\begin{array}{cc}
t_3(x^\mu,y) & \tilde t(x^\mu,y)\\
\tilde t(x^\mu,y)^\dagger & t_2(x^\mu,y)
\end{array}
\right),
\label{eq:frac_T}\\
S &=& \left(
\begin{array}{cc}
\sigma_3(y) \mathbf{1}_3 & 0\\
0 & \sigma_2(y) \mathbf{1}_2
\end{array}
\right)
+ \left(\begin{array}{cc}
s_3(x^\mu,y) & \tilde s(x^\mu,y)\\
\tilde s(x^\mu,y)^\dagger & s_2(x^\mu,y)
\end{array}
\right),
\label{eq:frac_S}
\ee
where the first terms on the right-hand sides are the background configurations.
The second terms stand for the small fluctuations where $t_3,s_3$ are 3 by 3, and $t_2,s_2$ are 2 by 2 Hermitian matrices, and
$\tilde t,\tilde s$ are 3 by 2 complex matrices.
Linearized equations of motion  can be cast into the following form: The diagonal parts are
of the form
\be
\left(-\partial_M\p^M - V_i\right) \bm{t}_i &=& 0,\qquad 
\bm{t}_i \equiv 
\left(
\begin{array}{c}
t_i\\
s_i
\end{array}
\right),
\label{eq:leom_t_scalar}
\ee
with $i = 3,2$.
Here, $\bm{t}_i$ is a two vector whose components are 3 by 3 matrices for $i=3$ and 2 by 2 matrices for $i=2$. 
The 2 by 2 symmetric matrix Schr\"odinger potential 
$V_i$ acts in the 2 component vector space $\bm{t}_i$.
The off-diagonal part has similar structure as
\be
\left(-\partial_M\partial^M  - \tilde V\right) \tilde{\bm{t}} &=& 0,\qquad 
\tilde{\bm{t}} \equiv 
\left(
\begin{array}{c}
\tilde t\\
\tilde s
\end{array}
\right),
\label{eq:leom_t_scalar2}
\ee
where $\tilde{\bm{t}}$ is a two vector whose components are 3 by 2 matrices, and
$\tilde V$ is  2 by 2 symmetric matrix again acting in the two-vector space $\tilde{\bm{t}}$.
The Schr\"odinger potentials are given as
\be
V_{i,11} &=& -2\lambda^2(v^2 - \tau_i^2 - \sigma_i^2) + 4 \lambda^2\tau_i^2,\label{eq:V_i,11}\\
V_{i,12} &=& V_{i,21} = 4\lambda^2 \tau_i\sigma_i,\\
V_{i,22} &=& -2\lambda^2(v^2 - \sigma_i^2 - \tau_i^2) + 4 \lambda^2\sigma_i^2 + \Omega^2,\\
\tilde V_{11} &=& \lambda^2(2\tau_3^2 +2\tau_3\tau_2 +2 \tau_2^2 + \sigma_3^2 + \sigma_2^2 - 2v^2) + \xi(\sigma_3-\sigma_2)^2,\\
\tilde V_{12} &=& \tilde V_{21} = \lambda^2(\tau_3 + \tau_2) (\sigma_3 + \sigma_2) - \xi (\tau_3 - \tau_2)(\sigma_3 - \sigma_2),
\label{eq:tV_11}\\
\tilde V_{22} &=& \lambda^2(2\sigma_3^2 +2\sigma_3\sigma_2 +2 \sigma_2^2 + \tau_3^2 + \tau_2^2 - 2v^2) + \xi(\tau_3-\tau_2)^2 + \Omega^2\,.
\ee

\begin{figure}
\begin{center}
\includegraphics[width=14cm]{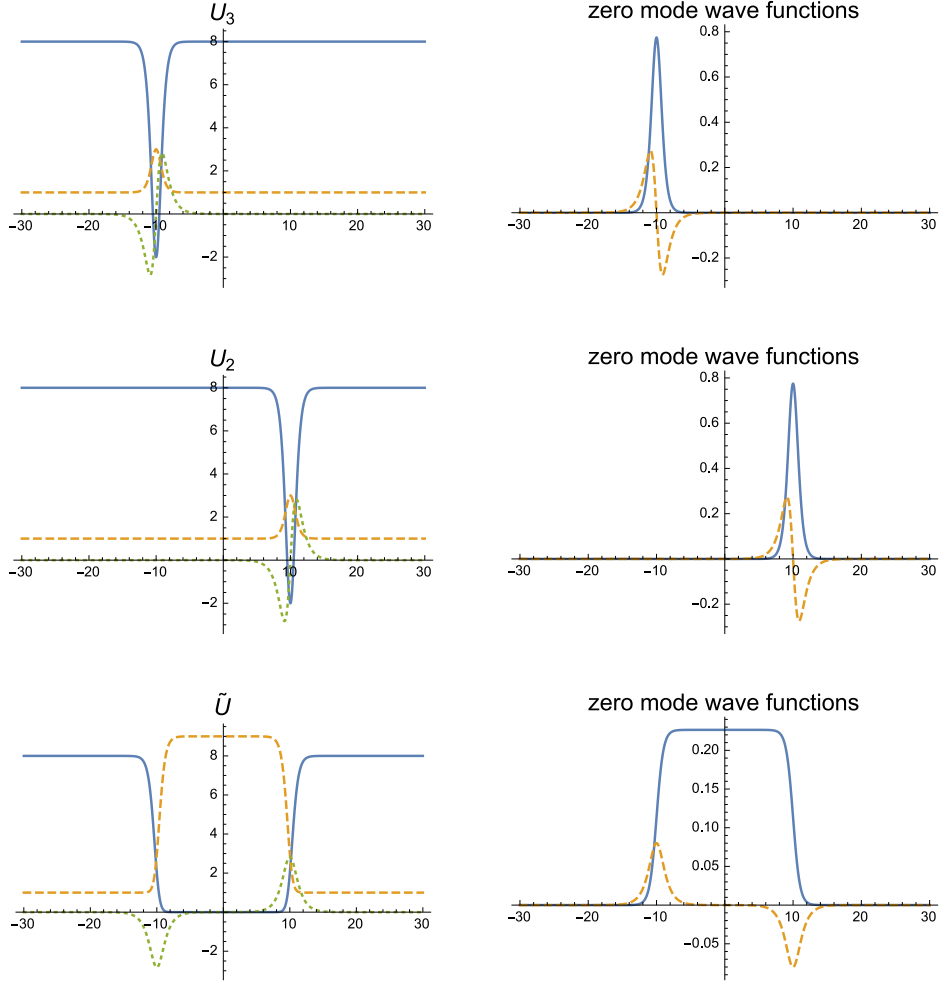}
\caption{The Schr\"odinger potentials $V_3$, $V_2$ and $\tilde V$ ($11$ component in solid line, $22$ component in dashed line,
and $12$ component in dotted line) are shown in the left panel in the top, middle and bottom row, respectively.
Corresponding zero mode wave functions $\mathbbm{u}_3^{(0)}$, $\mathbbm{u}_2^{(0)}$, and
$\tilde{\mathbbm{u}}^{(0)}$ (the upper component is in solid line and the lower component in the dashed line) are shown in the right panels.
${\cal Y}_3 = -10$ and ${\cal Y}_2 = 10$ with the model parameters $v=\sqrt{2}$, $\lambda=1$, $\Omega = 1$ and $\xi=1$.}
\label{fig:scalar_pot_zero_modes}
\end{center}
\end{figure}

Let us expand the fluctuation fields as
\be
\bm{t}_i(x^\mu,y) 
&=&
\sum_n
\left(
\begin{array}{c}
\eta_i^{(n)}(x^\mu) \mathbbm{u}_{i;1}^{(n)}(y)\\
\eta_i^{(n)}(x^\mu) \mathbbm{u}_{i;2}^{(n)}(y)
\end{array}
\right)
= \sum_n \eta_i^{(n)}(x^\mu) \otimes \mathbbm{u}_i^{(n)}(y),\\
\tilde{\bm{t}}(x^\mu,y) &=& 
\sum_n 
\left(
\begin{array}{c}
\tilde\eta_i^{(n)}(x^\mu) \tilde{\mathbbm{u}}_{i;1}^{(n)}(y)\\
\tilde\eta_i^{(n)}(x^\mu) \tilde{\mathbbm{u}}_{i;2}^{(n)}(y)
\end{array}
\right)
= \sum_n \tilde \eta^{(n)}(x^\mu) \otimes \tilde{\mathbbm{u}}^{(n)}(y),
\ee
where 
the basis $\mathbbm{u}_i^{(n)} = (\mathbbm{u}_{i;1}^{(n)},\ \mathbbm{u}_{i;2}^{(n)})^t$ and 
$\tilde{\mathbbm{u}}^{(n)}= (\tilde{\mathbbm{u}}_{i;1}^{(n)},\ \tilde{\mathbbm{u}}_{i;2}^{(n)})^t$
are two vectors whose components are scalar, and
the four-dimensional effective fields $\eta_3^{(n)}$ and $\eta_2^{(n)}$ are 3 by 3 and 2 by 2 
Hermitian matrices while $\tilde \eta^{(n)}$ is 3 by 2 complex matrix.
Note that the upper and lower components share the same four-dimensional effective fields $\eta_{i}^{(n)}$ and $\tilde{\eta}^{(n)}$.
The mass dimensions of the fields are $[\eta] = 1$ and $[\mathbbm{u}] = \frac{1}{2}$.
In order to figure out the spectrum, it is convenient to define the basis by
\be
\left(-\p_y^2 + V_i\right) \mathbbm{u}_i^{(n)} = m_{i,n}^2 \mathbbm{u}_i^{(n)},\qquad
\left(-\p_y^2 + \tilde V\right) \tilde{\mathbbm{u}}^{(n)} = \tilde m_{n}^2 \tilde{\mathbbm{u}}^{(n)}.
\ee
The wave functions of zero modes can explicitly be obtained as
\be
\mathbbm{u}_i^{(0)} &=& \frac{N_i}{\Omega}\, \p_y \left(
\begin{array}{c}
\tau_i\\
\sigma_i
\end{array}
\right),\label{eq:u0}\\
\tilde{\mathbbm{u}}^{(0)} &=& \tilde N \left(
\begin{array}{c}
\tau_3 - \tau_2\\
\sigma_3 - \sigma_2
\end{array}
\right),\label{eq:tildeu0}
\ee
where $N_i$ and $\tilde N$ stand for normalization constants whose mass dimensions are $[N_i] = [\tilde N] = -1$.
Fig.~\ref{fig:scalar_pot_zero_modes} shows the wave functions.
The former wave function $\mathbbm{u}^{(0)}_i$ is given by the $y$-derivative of the background solutions
in the diagonal components. This is expected because, for example, the zero modes in the
3 by 3 top-left diagonal small matrix is given by $\tau_3 + t_3 = v \tanh\Omega((y-{\cal Y}_3)\mathbf{1}_3 + Y_3 )$ 
and $\sigma_3 + s_3 = \bar v\,\sech\,\Omega((y-{\cal Y}_3)\mathbf{1}_3 + Y_3 )$ 
with $Y_3$ being arbitrary 3 by 3 Hermitian matrix. As usual, the zero mode wave function should be obtained 
by differentiating the solution in terms of the moduli parameters $Y_3$. Since $Y_3$ is a unique matrix appearing
in the solution, $Y_3$ derivative can be replaced by $y$ derivative. That is Eq.~(\ref{eq:u0}).
The zero mode of Eq.~(\ref{eq:tildeu0}) is obtained similarly.
These  fluctuations correspond to the NG bosons associated with $SU(5) \to SU(3) \times SU(2) \times
U(1)$ which we can see directly from infinitesimal transformation given by 
the off-diagonal elements of
Eq.~(\ref{eq:SU(5)_global_transf}).

Defining the inner product for two-component vectors of function of $y$ as
\be
(\mathbbm{u},\mathbbm{v}) \equiv
\int dy\, \mathbbm{u}^t~ \mathbbm{v}\,,
\label{eq:inner_prod}
\ee 
the normalization factors are 
determined by the condition 
$\left(\mathbbm{u}_i^{(0)},\mathbbm{u}_i^{(0)}\right) =  \left(\tilde{\mathbbm{u}}^{(0)},\tilde{\mathbbm{u}}^{(0)}\right) =1$. 
$\tilde N$ can be explicitly evaluated to give a function of the wall distance $L = |{\cal Y}_2 - {\cal Y}_3|$ as
\be
\frac{1}{\tilde N^2} 
= 4 L \left(\frac{\Omega^2}{\lambda^2}\frac{1}{\sinh L\Omega} + v^2 \tanh \frac{L\Omega}{2}\right) - \frac{4\Omega}{\lambda^2}\,.
\label{eq:tN}
\ee
\begin{figure}
\begin{center}
\includegraphics[width=8cm]{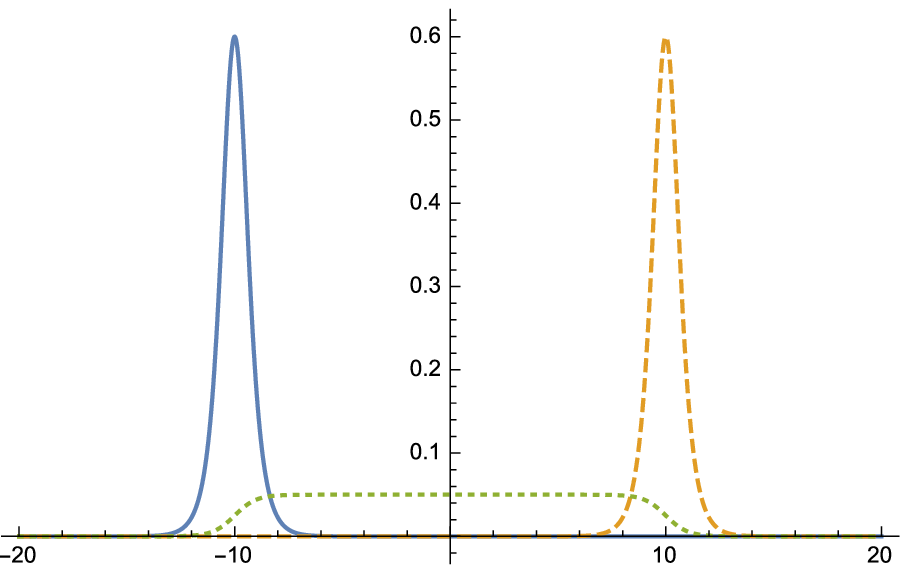}
\caption{The profiles of kinetic terms $\rho(\bm{t}_3^{(0)};y)$ 
(solid), $\rho(\bm{t}_2^{(0)};y)$ (dashed) and 
$\rho(\tilde{\bm{t}}^{(0)};y)$ (dotted) are shown for
${\cal Y}_3 = -10$ and ${\cal Y}_2 = 10$ with the model 
parameters $v=\sqrt{2}$, $\lambda=1$ and $\Omega = 1$,}
\label{fig:bg_scalar}
\end{center}
\end{figure}

To understand where the effective fields are localized, let us define the profiles of kinetic terms for zero modes as
\be
\rho(\bm{t}_i^{(0)};y) = \mathbbm{u}_i^{(0)t}~ \mathbbm{u}_i^{(0)},\qquad
\rho(\tilde{\bm t}^{(0)};y) = \tilde{\mathbbm{u}}^{(0)t}~ \tilde{\mathbbm{u}}^{(0)}\,,
\ee
with the mass dimension $[\rho] =1$ (compensating the mass dimension $-1$ from a $y$ integral). 
As illustrated by a typical example shown in Fig.~\ref{fig:bg_scalar},
the eight qNGs in $\eta_3^{(0)}(x^\mu)$ are localized on the left three coincident walls while
the three qNGs in $\eta_2^{(0)}(x^\mu)$ are localized on the right two coincident walls.
The profile of kinetic term for the translational NG mode is 
a linear combination 
$(\rho(\bm{t}^{(0)}_3;y) + \rho(\bm{t}^{(0)}_2;y))/2$, which   
has a support 
on both the left and right walls.
Finally, $\rho(\tilde{\bm t}^{(0)};y)$ 
provides the distribution for the twelve 
NGs  $\tilde \eta^{(0)}(x^\mu)$ associated with 
$SU(5) \to SU(3) \times SU(2) \times U(1)$ 
spreading between the left and right walls. 
The reason why $\tilde \eta^{(0)}$ localizes between walls is clear. It is because
the region between walls is asymptotically close to the vacuum $\left<3,2\right>$ where $SU(5)$ is partially broken.
They are called the non-Abelian cloud of the non-Abelian 
domain wall \cite{Eto:2008dm}.

\ \\\ \\
\noindent
{\bf The geometric Higgs mechanism}\\
As long as the $SU(5)$ is a global symmetry, 
the $25$ zero modes are all physical degrees of freedom.
However, once the gauge interaction is turned on, the $SU(5)$ 
becomes a local symmetry. Then the qNGs do 
remain as the physical zero modes whereas the 
NGs (except for the translational zero mode) will disappear 
from the physical spectra because they are absorbed into 
gauge bosons as their longitudinal components. 
Thus, the breaking pattern of the gauge symmetry is determined 
by the domain wall positions in the $y$ directions. 
By counting from the left-most wall, when the numbers of coincident 
walls is $(k_1,k_2,\cdots,k_n)$ with $\sum_{i=1}^n{k_i} = 5$, 
the gauge symmetry is broken as $SU(5) \to S[U(k_1) \times U(k_2) \times \cdots \times U(k_n)]$. 
One should note that the Higgs mechanism occurs locally 
since the would-be NG modes are localized between the split walls. 
This is the heart of the {\it geometric Higgs mechanism} which we are going to explain in detail in the subsequent section.
\\\ \\

Finally, let us make comments on massive modes.
In general, it is not easy to determine the massive modes
 because the linearized equations of motion \refer{eq:leom_t_scalar} 
 and \refer{eq:leom_t_scalar2} represent a coupled system of Schr\"odinger-like equations.
Nevertheless, some important information can be derived from 
the asymptotic values of the potentials 
\be
\lim_{y\to\pm \infty} V_3
= \lim_{y\to\pm \infty} V_2
= \lim_{y\to\pm \infty} \tilde V
= \left(
\begin{array}{cc}
4\lambda^2v^2 & 0\\
0 & \Omega^2
\end{array}
\right).
\label{eq:mass_gap_off_diagona_scalar}
\ee
Firstly, we see that all the off-diagonal components vanish (see dotted lines on the left panels of Fig.~\ref{fig:scalar_pot_zero_modes}). 
Therefore, the upper and lower components of $\mathbbm{u}_i^{(n)}$
and $\tilde{\mathbbm{u}}^{(n)}$ are asymptotically decoupled and become free. They interfere only near the domain walls.
Secondly, we see that there is a common mass gap between massless modes and continuum spectrum.
The mass gap is given by $\min\{2\lambda v,\Omega\} $. 
Massive modes will
be localized between the walls because they are bounded by the quasi-square well $\tilde V_{11}$ as is shown in the left-bottom panel (solid line)
of Fig.~\ref{fig:scalar_pot_zero_modes}.

\begin{figure}
\begin{center}
\includegraphics[width=15cm]{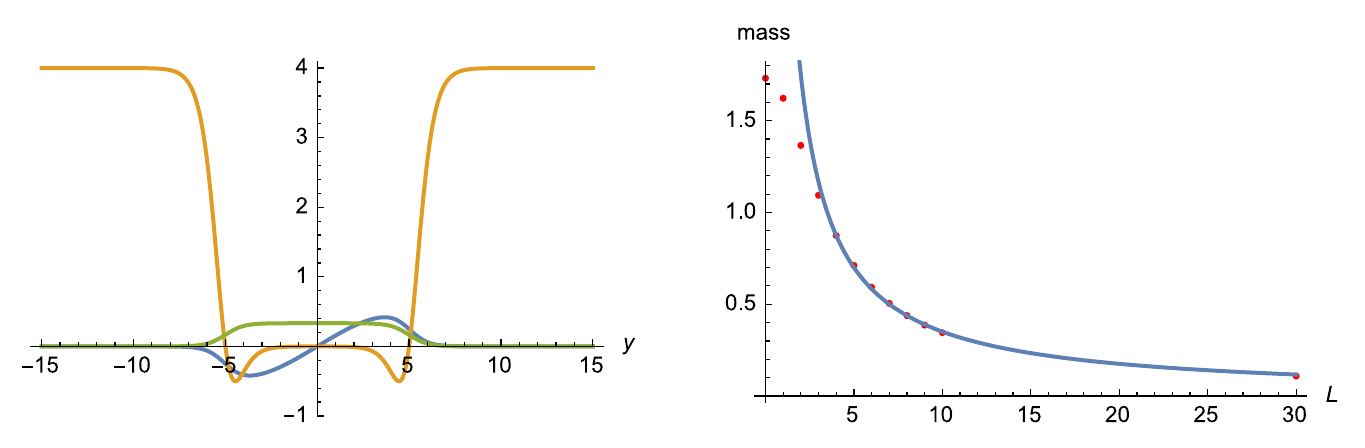}
\caption{Left: the Schr\"odinger potential $\tilde V_{11}$ and wave functions of the zero mode and the first excited mode in the off-diagonal component
are shown for
$L = 10$. Right: the masses of first excited modes are plotted in red dots. The solid curve is numerical fit by $3.5/L$. The parameters
are $\lambda = v = 1$.
}
\label{fig:mass_off_diagonal_reduced_model}
\end{center}
\end{figure}

In order to get a better insight,
let us further simplify the global $SU(5)$ model 
(\ref{eq:L_scalar_model}) and (\ref{eq:P_scalar_model}) by dropping the $S$ field.
Then, the simplified model is just an extension of $\phi^4$ type model with the adjoint scalar $T$ only.
The background wall solution is given by $T = v \tanh v\lambda(y-Y)$.
Let us consider fluctuations like in Eq.~(\ref{eq:frac_T}) with $\tau_i = v \tanh v\lambda (y-{\cal Y}_i)$ for the 3-2 splitting.
The Schr\"odinger equations for the fluctuations are obtained by just picking up the upper components of
Eqs.~(\ref{eq:leom_t_scalar}) and (\ref{eq:leom_t_scalar2}). The Schr\"odinger potential for the diagonal part is $V_{i,11}$ 
given in Eq.~(\ref{eq:V_i,11}) with $\sigma_i$ being replaced by $0$. 
This is nothing but the Schr\"odinger equation for linear 
fluctuations around a domain wall in ordinary $\phi^4$ model 
whose spectrum is well-known: the lowest modes are massless 
and the first excited modes have mass $\sqrt{3}\,\lambda v$. 
In our reduced model, these modes are the 3 by 3 and 2 by 2 
matrices in the adjoint representation of $SU(3)$ and $SU(2)$. 
Note that these modes are blind to whether the $SU(5)$ symmetry 
is global or local.
Similarly, the Schr\"odinger potential for the off-diagonal components is given by  $\tilde V_{11}$ in Eq.~(\ref{eq:tV_11}) with $\sigma_i \to 0$.
Since the existence of zero modes for the off-diagonal components is protected by symmetry, the upper elements 
of (\ref{eq:tildeu0}) remain as massless modes localized between the domain walls. 
On the other hand, we need a numerical computation to obtain excited modes since the 
Schr\"odinger equation cannot be solved analytically except for two extreme limits: zero separation limit $|{\cal Y}_3 - {\cal Y}_2| = 0$,
and infinite separation limit ${\cal Y}_2 \to \infty$ and ${\cal Y}_3 \to -\infty$, namely the vacuum $T = {\rm diag}(v,v,v,-v,-v)$.
In the former limit, $SU(5)$ symmetry is unbroken, and therefore both zero modes and excited modes form $SU(5)$ multiplets.
This means that $t_3$, $t_2$ and $\tilde t$ are all on an equal footing, so that mass of the first excited mode in $\tilde t$ should be $\sqrt{3}\,\lambda v$ as
that for $\phi^4$ kink. In the latter limit the off-diagonal components are the massless
NGs. Thus, for the finite separation $L$, the mass of the first excited mode $\tilde t$ 
is a continuous function, say $\tilde m(L)$, of the separation $L=|{\cal Y}_3-{\cal Y}_2|$, which asymptotically behaves as
$\tilde m \to \sqrt{3}\, \lambda v$ at $L\to 0$ and $\tilde m \to 0$ at $L \to \infty$.
Indeed, the Schr\"odinger potential $\tilde V_{11}$ at large $L$ is almost square well whose height is $4\lambda^2 v^2$ 
and width is $L$. Therefore, $\tilde m$ behaves as $1/L$ at the large $L$ limit, see Fig.~\ref{fig:mass_off_diagonal_reduced_model}. 
In short, the mass spectrum of the off-diagonal element 
for well separated domain walls starts
from zero and is followed by the massive modes of order $1/L$.

Thus we got an understanding that the off-diagonal components have a zero mode and light massive modes of order $1/L$ in the $SU(N)$ global model.
Whereas the zero mode will be eaten by the gauge fields,
one might anticipate the light massive modes appearing between the domain walls. 
But we emphasis that this is the case where $SU(5)$ is {\it global} symmetry.
As we will see in later sections, gauging $SU(5)$ will get rid of the off-diagonal zero modes and, at the same time,
it increases the masses of massive modes.

\section{Mass spectrum on domain walls in local $SU(N)$ model}\label{sec:threetwo}

Now, we come to main part of this work.
Our aim here is to determine the physical spectrum around the background domain walls (\ref{eq:back_true}) in the gauged model (\ref{eq:lagrangian}).
The case where all the $N$ domain walls are on top of each other has been intensively studied in Refs.~\cite{Ohta,Us1,Us2},
and the localization mechanism of massless $SU(N)$ gauge fields on the coincident walls is well understood.
In contrast, in this work, we will focus on  the case where some domain walls are separated from each other.
Especially, we will clarify how the massless $SU(N)$ gauge fields acquire non-zero masses, namely the geometric Higgs mechanism.

We continue to consider the $SU(5)$ model and the
$3$-$2$ split domain wall solution (\ref{eq:gensol3}) with $Y = {\rm diag}({\cal Y}_3,{\cal Y}_3,{\cal Y}_3,{\cal Y}_2,{\cal Y}_2)$, for its phenomenological significance.
Extension of our results to both generic number of walls and arbitrary configurations is straightforward. 

\subsection{Linearized equations of motion
}\label{sec:gauge2}

Let us derive linearized equations of motion for small fluctuations around the $3$-$2$ splitting background solution.
The fluctuations in the scalar fields $T,S$ are given in Eqs.~(\ref{eq:frac_T}) and (\ref{eq:frac_S}).
As this background breaks $SU(5)$ gauge symmetry down to standard model (SM) group $G_{\rm SM} = SU(3)\times SU(2)\times U(1)_Y$, 
let us parametrize the surviving symmetry transformations as
\begin{equation}\label{eq:gaugetrans2}
U_{5}  = \begin{pmatrix}
U_3 & 0 \\
0 & U_2 
\end{pmatrix} e^{i\frac{\alpha T_Y}{2}}\,,\qquad
T_Y = \left(
\begin{array}{cc}
-\frac{2}{3}\mathbf{1}_3 & 0\\
0 & \mathbf{1}_2
\end{array}
\right)\,,
\end{equation}
where $U_3 \in SU(3)$, $U_2 \in SU(2)$ and $\alpha \in U(1)_Y$. 
In the following, let us employ the axial gauge in $4+1$ dimensions 
\be
A_y =0\,.
\ee
There is a residual gauge transformation which depends only 
on $x^\mu$ coordinate. 
Further, let us separate diagonal and off-diagonal degrees of 
freedom in fluctuations of  the gauge fields as
\begin{equation}\label{eq:flucansatz2}
A_\mu = 0 + \begin{pmatrix}
a_{3\mu} & b_\mu \\
b_\mu^{\dagger} & a_{2\mu}
\end{pmatrix}
+ a_{1\mu} T_1\,,\qquad
T_1 = \sqrt{\frac{3}{5}}\left(
\begin{array}{cc}
\frac{1}{3}\mathbf{1}_3 & 0 \\
0 & - \frac{1}{2}\mathbf{1}_2
\end{array}
\right)\,,
\end{equation}
where $a_{3\mu}$ is a $3\times 3$ Hermitian traceless matrix 
and $a_{2\mu}$ is a $2\times 2$ Hermitian traceless matrix, 
while $b_\mu$ is a $3\times 2$ complex matrix.
The gauge fields $a_{3\mu}$, $a_{2\mu}$, and $a_{1\mu}$ 
for $SU(3)$, $SU(2)$ and $U(1)_Y$ transform under the SM gauge 
group as  
\begin{align}
a_{3\mu} & \to U_3  a_{3\mu}U_3^{\dagger} +i \partial_\mu U_3 U_3^{\dagger}\,, \\
a_{2\mu} & \to U_2  a_{2\mu}U_2^{\dagger} +i \partial_\mu U_2 U_2^{\dagger}\,, \\
a_{1\mu} & \to a_{1\mu} + \sqrt{\frac{5}{3}} \p_\mu\alpha. 
\end{align}
On the other hand, the $b_\mu$ field transforms as
\begin{equation}
b_\mu \to U_3 b_\mu U_2^{\dagger}e^{-\frac{5 i\alpha}{6}}\,.
\end{equation}

To investigate the spectrum, we need to write down the linearized equations of motion for each component of \refer{eq:frac_T},
\refer{eq:frac_S} and \refer{eq:flucansatz2}. 
Plugging these into equations of motion \refer{eq:fullEOM1} -- \refer{eq:fullEOM3}, we 
end up with
\begin{align}
\partial_M\Bigl(\sigma_\alpha^2f_\alpha^{MN}\Bigr) & = 0\,, \label{eq:f_split1}\\
\left(-\partial_M\p^M - V_i\right) \bm{t}_i &= 0\,, \label{eq:f_split3}\\
a\p_M\left(\sigma_+^2 C^{MN}\right) &= - \frac{i}{\tilde N}  \tilde{\mathbbm{u}}^{(0)t}
\left(\overleftrightarrow{\p}^N \tilde{\bm{t}} - \frac{i}{\tilde N} b^N  \otimes \tilde{\mathbbm{u}}^{(0)}\right)\,,\label{eq:f_split4}\\
\left(-\partial_M\partial^M  - \tilde V\right) \tilde{\bm{t}} &= - \frac{i}{\tilde N}\left(\p_Mb^M\right)  \tilde{\mathbbm{u}}^{(0)}
- \frac{2i}{\tilde N}b^M \otimes \p_M\tilde{\mathbbm{u}}^{(0)}\,,\label{eq:f_split5}
\end{align}
where no sum is taken for $\alpha = 1,2,3$ and $i=3,2$ in 
Eqs.~(\ref{eq:f_split1})--(\ref{eq:f_split3}).  
The linearized field strength is defined as usual by 
$f_\alpha^{MN} = \p^M  a_\alpha^N - \p^N a_\alpha^M$, and
$\sigma_i$ ($i=2,3)$ is defined in (\ref{eq:wall22}).
In addition, we have introduced 
\be
\sigma_1 = \sqrt{\frac{2 \sigma_3^2 + 3 \sigma_2^2}{5}}\,,\quad
\sigma_+ = \sqrt{\sigma_3^2 + \sigma_2^2}\,,\quad
C^{MN} = \p^M b^N - \p^N b^M\,,
\ee 
and $\tilde{\mathbbm{u}}^{(0)}$ is given in Eq.~(\ref{eq:tildeu0}).

\subsection{Diagonal components}

First, we find that the fluctuations $a_\alpha^\mu$ ($\mu=0,1,2,3$, $\alpha=1,2,3$) and $\bm{t}_i$ ($i=3,2$)
in the diagonal parts are decoupled from the other fields.
Especially, Eq.~(\ref{eq:f_split3}) for $\bm{t}_i$ is exactly 
the same as Eq.~(\ref{eq:leom_t_scalar}).
Therefore, we have $9$ and $4$ zero modes in $\bm{t}_3$ and $\bm{t}_2$ whose wave functions have been determined as
$\mathbbm{u}_{3}^{(0)}$ and $\mathbbm{u}_{2}^{(0)}$ given in Eq.~(\ref{eq:u0}).

Let us next investigate  spectrum for the unbroken parts of the gauge fields given in Eq.~(\ref{eq:f_split1}).
The $N=y$ component in the axial gauge ($b_y = 0$) is
\be
\sigma_\alpha^2 \p_y\p_\mu a_\alpha^\mu = 0\,,\qquad\text{(no sum for $\alpha$)},
\ee
and the $N=\nu$ component is
\be
\sigma_\alpha^2 \p_\mu f_\alpha^{\mu\nu} + \p_y (\sigma_\alpha^2 \p^y  a_\alpha^\nu) = 0\,,\qquad\text{(no sum for $\alpha$)}.
\ee
We can decompose the gauge field into divergence-free and divergence components as
\be
a_\alpha^\mu =  a_{\alpha,{\rm df}}^{\mu} +  a_{\alpha,{\rm d}}^{\mu}\,\quad  
a_{\alpha,{\rm df}}^{\mu} = ({\cal P}_{\rm df})^\mu{}_\nu a_\alpha^\nu\,,\quad  
a_{\alpha,{\rm d}}^{\mu} = ({\cal P}_{\rm d})^\mu{}_\nu a_\alpha^\nu\,,
\ee
where we introduced the projection operators 
\be
({\cal P}_{\rm d})^\mu{}_\nu = \frac{\partial^\mu \partial_\nu}{\partial^2}\,,\qquad
({\cal P}_{\rm df})^\mu{}_\nu = \delta^\mu_\nu-\frac{\partial^\mu \partial_\nu}{\partial^2}\,,
\label{eq:ddf}
\ee
with the four-dimensional Laplacian  $\p^2 = \p_\mu\p^\mu$.
They satisfy the following identities 
\be
\p_\mu a_{\alpha,{\rm df}}^\mu = 0\,,\quad
a_{\alpha,{\rm d}}^{\mu} = \frac{1}{\p^2}\p^\mu F_\alpha,\quad F_\alpha = \p_\nu a_\alpha^\nu\,.
\ee
The $N=y$ equation tells us that $\p_y F_\alpha = 0$, so that we have $F_\alpha = F_\alpha(x^\mu)$ which
can be gauged away by using $SU(3) \times SU(2) \times U(1)_Y$ gauge transformation.
Then the $N=\nu$ component reads 
\be
\sigma_\alpha^2 \left(\p^2 - \p_y^2\right) a_{\alpha,{\rm df}}^{\nu} - (\p_y \sigma_\alpha^2)\p_y a_{\alpha,{\rm df}}^{\nu} = 0\,,\qquad\text{(no sum for $\alpha$)}.
\ee
To find the spectrum, let us expand the divergence-free component as
\be \label{eq:wavea}
a_{\alpha,{\rm df}}^{\nu} = \sum_n w_{\alpha\nu}^{(n)}(x^\mu)f_{a,\alpha}^{(n)}(y) 
\equiv \sum_n w_{\alpha\nu}^{(n)}(x^\mu) \frac{v_\alpha^{(n)}(y)}{\sigma_\alpha(y)},
\ee
where $w_{\alpha\nu}^{(n)}(x^\mu)$ is the four-dimensional gauge fields (matrix) and $v_\alpha^{(n)}(y)/\sigma_\alpha(y)$ 
is its wave function (one component). 
The mass dimensions are given by $[w_{\alpha\mu}^{(n)}] = 1$ and  $[v_\alpha^{(n)}] = \frac{3}{2}$.
The basis of expansion is defined by the Schr\"odinger equation
\be
\left(-\p_y^2 + V_\alpha\right)v_\alpha^{(n)} = \mu_{\alpha,n}^2 v_\alpha^{(n)},\quad
V_\alpha \equiv \frac{1}{\sigma_\alpha}\p_y^2 \sigma_\alpha = \Omega^2\left(1-2\,\sech^2\Omega(y-{\cal Y}_\alpha)\right),
\label{eq:pot20}
\ee
where no sum is taken for $\alpha$.

\begin{figure}
\begin{center}
\includegraphics[width=8cm]{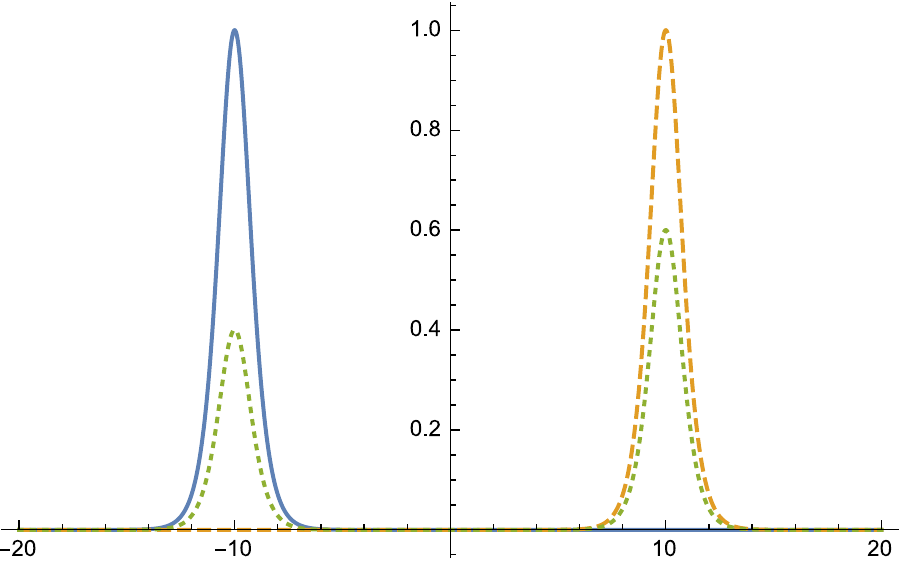}
\caption{The profiles of kinetic terms for $\rho(w_{3\mu}^{(0)};y)$ (solid), $\rho(w_{2\mu}^{(0)};y)$ (dashed) and $\rho(w_{1\mu}^{(0)};y)$ (dotted) are shown for
${\cal Y}_3 = -10$ and ${\cal Y}_2 = 10$ with the model parameters $v=\sqrt{2}$, $\lambda=1$ and $\Omega = 1$,}
\label{fig:diag_gauge_wavefunc}
\end{center}
\end{figure}

This Schr\"odinger-type problem can be cast into the following form
\be
H_\alpha v_\alpha^{(n)} = \mu_{\alpha,n}^2 v_\alpha^{(n)}\,,\quad
H_\alpha = Q_\alpha^\dagger Q_\alpha\,,
\ee
where we define
\be
Q_\alpha = \p_y - \p_y\log \sigma_\alpha\,,\quad
Q_\alpha^\dagger = -\p_y - \p_y\log \sigma_\alpha\,.
\ee
There are two benefits for this expression. First, the Hamiltonian is manifestly positive definite, so that we can be sure that no tachyonic modes exist in the spectrum.
Second, 
the zero mode can be easily found by solving $Q v_\alpha^{(0)} = 0$, which gives
\be
v_\alpha^{(0)}(y) = {\cal N}_\alpha \,\sigma_\alpha(y),
\label{eq:diagonal_gauge_zero}
\ee
where ${\cal N}_\alpha$ stands for normalization constant of the mass dimension $[{\cal N}_\alpha] = 0$.
The normalization factor ${\cal N}_\alpha$ is fixed as ${\cal N}_\alpha=1$ to have properly normalized field strength $w_\alpha^{(0)\mu\nu}\equiv 
\partial^\mu w_\alpha^{(0)\nu}-\partial^\nu w_\alpha^{(0)\mu} +i \comm{w_\alpha^{(0)\mu}}{w_\alpha^{(0)\nu}}$.
Note that the zero mode wave functions are flat ${\cal N}_\alpha v_\alpha^{(0)}/\sigma_\alpha = 1$, nevertheless, the massless effective gauge fields are localized on the walls thanks to
the Ohta-Sakai gauge kinetic function (\ref{eq:ext_OS_factor}).
The profile of kinetic terms for the zero mode $w_\alpha^{(0)\mu}$  can be read as
\be
\rho(w_{\alpha\mu}^{(0)};y) = a\sigma_\alpha^2  \times \left(\frac{v_\alpha^{(0)}}{\sigma_\alpha}\right)^2 = a(v_\alpha^{(0)})^2 = a\sigma_\alpha^2
\,,\qquad\text{(no sum for $\alpha$)},
\ee
where the factor $a\sigma_\alpha^2$ reflects the 
gauge kinetic function (\ref{eq:ext_OS_factor}). 
The mass dimension is $[\rho(w_{\alpha\mu}^{(0)};y)] = 1$.
Fig.~\ref{fig:diag_gauge_wavefunc} shows a typical profiles of 
the massless gauge fields. 
It clearly demonstrates that $SU(3)$ gauge fields localize 
on the left three coincident walls and $SU(2)$ gauge fields 
are trapped by the right two coincident walls. $U(1)_Y$ gauge 
fields have supports both on left and right walls.

In general, the Schr\"odinger equation with the Hamiltonian 
\begin{equation}
H = -\partial_y^2 - \frac{c}{\cosh^2(y)}
\end{equation}
has a finite number of discrete boundstates. 
Their energies are given by the textbook formula
\begin{equation}\label{eq:textbook}
E_n = -\left(\sqrt{c+\frac{1}{4}}-\frac{1}{2}-n\right)^2\,,
\end{equation} 
where $n$ takes nonnegative integer values 
starting from 0 up to the number, 
for which the expression in the parenthesis is still positive. 
Given this fact, it is easy to see that for the potential 
\refer{eq:pot20} there is only the zero mode as a bound state.
No other massive discrete bound states exist, 
while the mass gap between the zero mode and the continuum modes 
is $\Omega$ which is of order $M_5$.

The effective gauge coupling constants for the effective 
$SU(3) \times SU(2) \times U(1)_Y$ gauge group can be read as follows.
Let us first decompose the gauge kinetic term (\ref{eq:ohta}) with
the fluctuations $a_{i\mu}$ and $b_\mu = 0$ in Eq.~(\ref{eq:flucansatz2}),
\be
\Tr\left[S^2 G_{MN}G^{MN}\right] =  \sigma_3^2 \Tr\left[f_{3\mu\nu}f_3^{\mu\nu}\right]
+  \sigma_2^2 \Tr \left[f_{2\mu\nu}f_2^{\mu\nu}\right]
+  \left(\frac{\sigma_3^2}{5}+ \frac{3\sigma_2^2}{10}\right)f_{1\mu\nu}f_1^{\mu\nu}\,.
\ee
Integrating this over $y$, we find the effective gauge 
couplings as
\be
-\frac{1}{2g_3^2} \Tr\left[f_{3\mu\nu}f_3^{\mu\nu}\right]
-\frac{1}{2g_2^2} \Tr \left[f_{2\mu\nu}f_2^{\mu\nu}\right]
-\frac{1}{4g_1^2} f_{1\mu\nu}f_1^{\mu\nu}\,,
\ee
with
\begin{equation}
\label{eq:4dim_g}
\frac{1}{g_3^2}  =  \frac{1}{g_2^2} 
= \frac{1}{g_1^2} 
= \frac{4a\bar v^2}{\Omega} \equiv \frac{1}{g_5^2}\,.
\end{equation}
These are the dimensionless gauge coupling constants in $3+1$ dimensions.
We see that the effective gauge couplings $g_2$ and $g_3$ are given by parameters of the model and that they are equal to each other. 
the $U(1)_Y$ coupling $e_Y$ is given by $e_Y = - \sqrt{\frac{3}{20}} g_1$, and 
is related to $g_2$ and $g_3$ as 
\be
\frac{1}{4e_Y^2} =\frac{2}{3 g_3^2}+\frac{1}{g_2^2} =  \frac{20a \bar v^2}{3\Omega}\,.
\ee
These relations are identical to the standard $SU(5)$ GUT 
scenario. Hence the prediction of Weinberg angle at the GUT 
scale is also the same as the standard $SU(5)$ GUT: 
 $\theta_W = \arctan (2e_Y/g_2)$ is given as $\tan^2\theta_W = 3/5$. 
This purely group-theoretical result arises because of the 
identical profile of position dependent gauge couplings for these 
gauge groups in our simple model. 
However, we can obtain different profiles for different 
gauge coupling function and a deviation from the standard $SU(5)$ 
GUT, if we consider models with more complex structure.

\subsection{The geometric Higgs mechanism}\label{sec:geomhiggs}

Let us next investigate the off-diagonal parts in 
Eqs.~(\ref{eq:f_split4}) and (\ref{eq:f_split5}). 
These are coupled equations for the fluctuations $\tilde{\bm{t}}$ 
and $b^M$.
As we have shown in Eq.~(\ref{eq:tildeu0}), there is 
a zero mode 
$\tilde \eta^{(0)}(x^\mu) \otimes \tilde{\mathbbm{u}}^{(0)}(y)$ 
in $\tilde{\bm{t}}$ before coupling to the $SU(5)$ 
gauge field. We are going to show that 
this zero mode disappears from the physical spectrum
once the gauge interaction is turned on. 
This is a manifestation of the geometric Higgs mechanism. 
We continue to use the axial gauge $b_y = 0$.

Let us separate the zero mode and define fields $\bar{\bm t}$ 
containing only massive modes as
\be
\bar{\bm t}(x^\mu,y) = \tilde{\bm t}(x^\mu,y) 
- \tilde \eta^{(0)}(x^\mu) \otimes \tilde{\mathbbm{u}}^{(0)}(y),
\label{eq:zero_subtraction}
\ee
where we have defined a $3$ by $2$ matrix
\be
\tilde \eta^{(0)}(x^\mu) = \left( \tilde{\mathbbm{u}}^{(0)}(y), \tilde{\bm t}(x^\mu,y)\right)\,.
\ee
Note that the inner product should be taken by means of 
Eq.~(\ref{eq:inner_prod}), and remember that 
the four-dimensional fields $\tilde \eta^{(0)}(x^\mu)$ is 3 
by 2 matrix.
Thus, $\bar{\bm t}$ includes only massive modes orthogonal 
to $\tilde{\mathbbm{u}}^{(0)}$.
Let us rewrite Eqs.~(\ref{eq:f_split4}) and (\ref{eq:f_split5})  
by using $\bar{\bm t}$.
The $N=\nu$ and $N = y$ components of (\ref{eq:f_split4}) 
are of the form
\be
a \sigma_+^2 \p_\mu \tilde C^{\mu\nu} + a \p_y 
\left(\sigma_+^2 \p^y \tilde b^\nu\right) &=& 
- \frac{i}{\tilde N} \tilde{\mathbbm{u}}^{(0)t}\,
\p^\nu \bar{\bm{t}} - \frac{\tilde{\mathbbm{u}}^{(0)t}\,
\tilde{\mathbbm{u}}^{(0)}}{\tilde N^2}  \tilde b^\nu\,,
\label{eq:off_g1}\\
a \sigma_+^2 \p^y \p_\mu \tilde b^\mu &=& \frac{i}{\tilde N} 
\tilde{\mathbbm{u}}^{(0)t}\, \overleftrightarrow{\p^y} \bar{\bm t}\,,
\label{eq:off_g2}
\ee
where we have defined 
\be
\tilde b^\nu(x^\mu,y) &=&  b^\nu(x^\mu,y) 
+ i \tilde N \p^\nu \tilde \eta^{(0)}(x^\mu)\,,
\label{eq:zero_subtraction2}\\
\tilde C^{\mu\nu} &=& \p^\mu \tilde b^\nu - \p^\nu \tilde b^\mu\, .
\ee 
Eq.~(\ref{eq:f_split5}) is also rewritten as
\be
- \left(\p^2 + \p_y^2 - \tilde V\right) \bar{\bm t} 
= - \frac{i}{\tilde N} \left(\p_\mu \tilde b^\mu\right) 
\otimes \tilde{\mathbbm{u}}^{(0)}\,.
\label{eq:off_s}
\ee
Now, we are left with Eqs.~(\ref{eq:off_g1}), (\ref{eq:off_g2}) 
and (\ref{eq:off_s}), and we should note that 
the off-diagonal scalar zero mode $\tilde \eta^{(0)}(x^\mu)$ 
does not appear alone but is hidden in $\tilde b^\nu$.
This is nothing but what happens for the standard Higgs 
mechanism: a massless vector field eats a scalar NG mode
and acquires a mass. 
Indeed, one realizes that Eq.~(\ref{eq:zero_subtraction}) 
and (\ref{eq:zero_subtraction2}) are nothing but the residual gauge 
transformation $U_5 = U_5(x^\mu) \in SU(5)$ in the axial gauge $A_y = 0$.
We perform the same infinitesimal $SU(5)$ transformation as 
Eq.~(\ref{eq:SU(5)_global_transf}). The only difference here
is that the transformation is the gauge transformation. 
Transforming $T$ and $S$ given in Eqs.~(\ref{eq:frac_T}) and 
(\ref{eq:frac_S}) 
by $U_5$  given in Eq.~(\ref{eq:infinitesimal_SU5)}) with 
a $3$ by $2$ matrix $\tilde \alpha(x^\mu)$ of local 
transformation parameter, we find 
\be
\tilde{\bm{t}}(y,x^\mu) \to \tilde{\bm{t}}'(y,x^\mu) 
=\tilde{\bm{t}}(y,x^\mu)- \frac{i}{\tilde N} \tilde \alpha(x^\mu) 
\otimes \tilde{\mathbbm{u}}^{(0)}(y). 
\label{eq:unitary_gauge_t}
\ee
Similarly, the same infinitesimal gauge transformation of 
the gauge field given in Eq.~(\ref{eq:flucansatz2}) gives
\be
b_\mu(y,x^\mu) \to 
b'_\mu(y,x^\mu)= b_\mu(y,x^\mu) - \p_\mu \tilde \alpha(x^\mu). 
\label{eq:unitary_gauge_b}
\ee
It is easy to see that gauge transformed $\tilde{\bm t}'$ 
in Eq.~(\ref{eq:unitary_gauge_t}) 
and $b_\mu'$ in Eq.~(\ref{eq:unitary_gauge_b}) 
can be identified as $\bar{\bm t}$ in Eq.~(\ref{eq:zero_subtraction}) 
and $\tilde{b}_\mu$ in Eq.~(\ref{eq:zero_subtraction2}), 
by choosing the gauge transformation parameter as 
$\tilde \alpha(x^\mu) = -i \tilde N \tilde\eta^{(0)}(x^\mu)$. 
We call this choice of gauge 
as the unitary gauge for the geometric Higgs mechanism.

Note also that Eq.~(\ref{eq:off_g2}) is redundant because it 
can be derived by a combination of Eq.~(\ref{eq:off_g1}) 
(after operating $\p_\nu$) and Eq.~(\ref{eq:off_s}) (after multiplying 
$\tilde{\mathbbm{u}}^{(0)t}$ from left).  
Therefore, the spectrum is determined by Eqs.~(\ref{eq:off_g1}) 
and (\ref{eq:off_s}).

Let us decompose Eqs.~(\ref{eq:off_g1}) and (\ref{eq:off_s})
into divergence and divergence-free parts by applying the projection operators given in Eq.~(\ref{eq:ddf}) as
\be
\tilde b^\mu = \tilde b_{\rm d}^\mu + \tilde b_{\rm df}^\mu\,,\qquad
\tilde b_{\rm d}^\mu \equiv ({\cal P}_{\rm d})^\mu{}_\nu\tilde b^\nu\,,\qquad
\tilde b_{\rm df}^\mu \equiv ({\cal P}_{\rm df})^\mu{}_\nu\tilde b^\nu\,,
\ee
with $\p_\mu \tilde b_{\rm df}^\mu = 0$ due to $({\cal P}_{\rm df})^\mu{}_\nu \p^\nu = 0$.
Now, Eqs.~(\ref{eq:off_g1}) and (\ref{eq:off_s}) are written as
\be
a \sigma_+^2 \p^2 \tilde b_{\rm df}^\nu + a \p_y\left(\sigma_+^2\p^y \tilde b_{\rm df}^\nu\right)
&=& - \frac{\tilde{\mathbbm{u}}^{(0)t}\,\tilde{\mathbbm{u}}^{(0)}}{\tilde N^2}  \tilde b_{\rm df}^\nu\,,
\label{eq:leom_df}\\
a \p_y\left(\sigma_+^2\p^y\tilde b_{\rm d}^\mu\right) 
&=& - \frac{i}{\tilde N} \tilde{\mathbbm{u}}^{(0)t}\left(\p^\mu \bar{\bm{t}} - \frac{i}{\tilde N} \tilde b_{\rm d}^\mu \otimes \tilde{\mathbbm{u}}^{(0)}\right)\,,
\label{eq:leom_f}\\
- \left(\p_y^2 - \tilde V\right) \bar{\bm t} 
&=& \p_\mu\left(\p^\mu \bar{\bm t} - \frac{i}{\tilde N} \tilde b_{\rm d}^\mu \otimes \tilde{\mathbbm{u}}^{(0)}\right)\,.
\label{eq:leom_sc}
\ee

Benefit of this decomposition is that the divergence-free part $\tilde b_{\rm df}^\mu$ is decoupled from
the other fields.
For the time being, we will concentrate on Eq.~(\ref{eq:leom_df}) and find the spectrum of $\tilde b_{\rm df}^\mu$.
To this end, let us expand 
\be
\tilde b_{\rm df}^\mu(x^\nu,y) = \sum_n \beta^{(n)\mu}(x^\nu) \frac{\gamma^{(n)}(y)}{\sigma_+(y)}\,.
\label{eq:tb_expansion}
\ee
where $\beta_\mu^{(n)}$ is 3 by 2 complex matrix satisfying the divergence-free condition $\p^\mu \beta_\mu^{(n)} = 0$.
The mass dimensions are $[\beta_\mu^{(n)}] = 1$ and $[\gamma^{(n)}] = \frac{3}{2}$.
Plugging this into Eq.~(\ref{eq:leom_df}), we are lead to
\begin{align}
& \left(\p^2 + k_n^2\right)\beta^{(n)\nu} = 0\,,
\label{eq:off_gmu}\\
&\left(-\p_y^2 + {\cal V}(y)\right)\gamma^{(n)} = \tilde \mu_n^2 \gamma^{(n)}\,,\\
&{\cal V}(y) = \frac{1}{\sigma_+}\p_y^2 \sigma_+ 
+ \frac{1}{\tilde N^2} \frac{\tilde{\mathbbm{u}}^{(0)t}\, \tilde{\mathbbm{u}}^{(0)}}{a\sigma_+^2}\,.
\label{eq:eff_pot}
\end{align}
Note that the Schr\"odinger equation can be written in the following form
\be
\tilde H \gamma^{(n)} = \tilde \mu_n^2 \gamma^{(n)}\,,
\label{eq:H}
\ee
with the Hamiltonian
\be
\tilde H = \tilde Q^\dagger \tilde Q 
+ \frac{1}{\tilde N^2} \frac{\tilde{\mathbbm{u}}^{(0)t}\, \tilde{\mathbbm{u}}^{(0)}}{a\sigma_+^2}\,, \label{eq:ham}
\ee
where we have defined differential operators
\be
\tilde Q = \p_y - \p_y\log \sigma_+,\qquad
\tilde Q^\dagger = -\p_y - \p_y\log \sigma_+.
\label{eq:QQ}
\ee
Note that the second term of the Hamiltonian $\tilde H$ is positive everywhere if ${\cal Y}_3 \neq {\cal Y}_2$.
Only when ${\cal Y}_3 = {\cal Y}_2$ it vanishes and Hamiltonian becomes $\tilde H = \tilde Q^\dagger \tilde Q$. In this coincident wall limit, 
there exists a zero mode which satisfies $\tilde Q \gamma^{(0)} = 0$. It is easily solved as
\be
\gamma^{(0)} = N_+ \sigma_+\,,\qquad
N_+ = 1\,.
\label{eq:zero_off_b}
\ee

\begin{figure}
\begin{center}
\includegraphics[width=16cm]{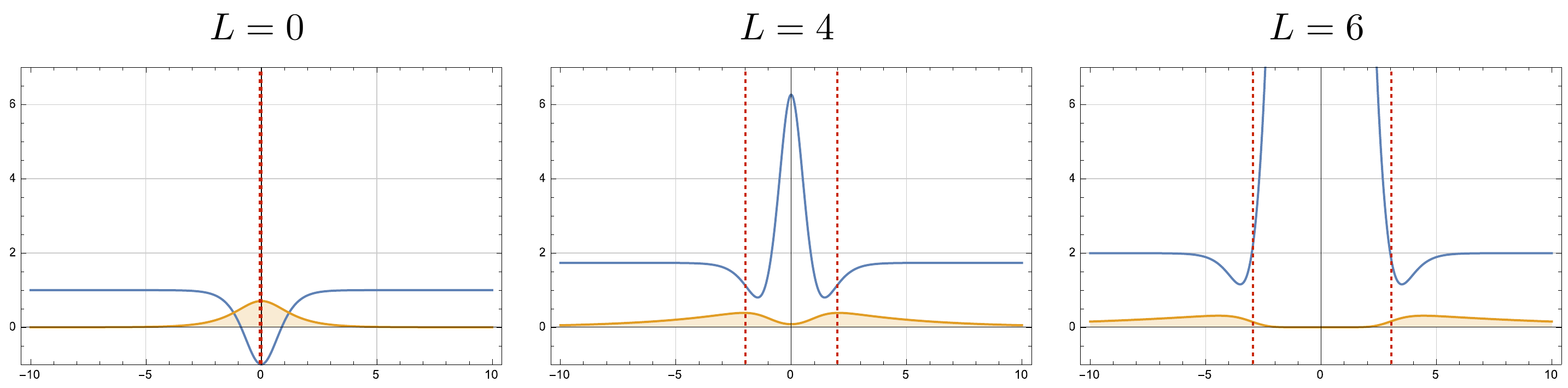}
\caption{The Schr\"odinger  potential ${\cal V}$ (blue solid) and the wave function (orange solid filling) of the lightest mode $\gamma^{(0)}$ for $L = 0$ , 
$L = 4$  and $L = 6$  are shown.
The model parameters are chosen as $a=1$, $v=\sqrt{2}$, $\lambda=1$ and $\Omega = 1$. The red dashed lines show  positions of the domain walls.}
\label{fig:sch_pot_off}
\end{center}
\end{figure}
The zero mode (\ref{eq:zero_off_b}) should exist because $SU(5)$ gauge symmetry is fully unbroken in the coincident wall limit where 
not only the diagonal components but also off-diagonal components of the gauge fields are massless.
Since $\tilde Q^\dagger \tilde Q$ is positive definite, the zero eigenvalue is minimum among all other eigenvalues.
When the 3-2 splitting occurs, no matter how small the separation is, the second term of $H$ has a positive contribution to $\tilde Q^\dagger \tilde Q$.
Therefore, the minimum of the spectrum for Eq.~(\ref{eq:H}) is positive whenever the walls split. 
Thus, we conclude $\tilde \mu_n^2 > 0$ for the 3-2 splitting background. 
Let us write the lowest mass field as $\beta_\mu = n_\mu e^{i k^\nu x_\nu}$ with $k^2 = \tilde \mu_0^2$.
The transverse condition $\p_\mu \beta^\mu = 0$ implies $k_\mu n^\mu = 0$. Since $k^2 = \tilde \mu_0^2 > 0$,
there are three orthonormal vectors $n_a^\mu$ ($a=1,2,3)$. Namely, the vector field $\tilde b^\mu_{\rm df}$ orthonormal to $k^\mu$ is 
massive  with three physical degrees of freedom (2 transverse and 1 longitudinal).
This is evidently due to the  geometric Higgs mechanism.

\begin{figure}
\begin{center}
\includegraphics[width=10cm]{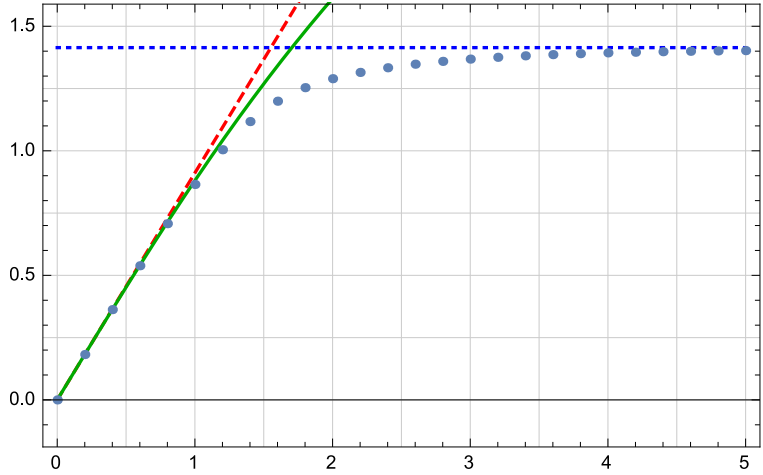}
\caption{
The  mass $\tilde \mu_0(L)$ of the lowest discrete massive 
state of the off-diagonal divergence-free vector field 
$\tilde b_{\rm df}^\mu$ as a function of $L = |{\cal Y}_3 - {\cal Y}_2|$.
The dots are numerically obtained, the green-solid curve is analytic approximation given in Eq.~(\ref{eq:mass_small_L_0}),
and the red-dashed line is a linear approximation with $k_0(0) = 0$ and $k_0(0.1) = 0.182252$.
The model parameters are chosen as $a=1$, $v=\sqrt{2}$, $\lambda=1$ and $\Omega = 1$.
The blue-dotted line shows the mass gap $\sqrt{\Omega^2 + 1/a} = \sqrt{2}$.}
\label{fig:mass_off_diagonal}
\end{center}
\end{figure}

Let us obtain the first massive mode. 
The Schr\"odinger potential ${\cal V}$ for $L = |{\cal Y}_2-{\cal Y}_3| = 0,4,6$ cases are
shown in Fig.~\ref{fig:sch_pot_off}.
In the limit with $\Omega L \ll 1$ where the separation is very small, the second term of ${\cal V}$ can be treated as
a perturbation to $L = 0$ case. 
Since there exists a localized  zero mode $\gamma^{(0)} =  \sigma_+$ in the $L=0$ limit, we expect 
the bound state remains as massive state as long as 
$\Omega L \ll 1$. 
The mass shift is estimated as
\be
\tilde \mu_0(L)^2 &=& \frac{\bigg<\gamma^{(0)}\bigg|
\frac{1}{\tilde N^2} \frac{\tilde{\mathbbm{u}}^{(0)t} \tilde{\mathbbm{u}}^{(0)}}{a\sigma_+^2} \bigg|\gamma^{(0)}\bigg>}{
\big<\gamma^{(0)}\big|\gamma^{(0)}\big>}
= \frac{1}{a \tilde N^2 \big<\gamma^{(0)}\big|\gamma^{(0)}\big>}  \nonumber\\
&=&\frac{\Omega}{4a\bar v^2}
\left[
4 L \left(\frac{\Omega^2}{\lambda^2}\frac{1}{\sinh L\Omega} + v^2 \tanh \frac{L\Omega}{2}\right) - \frac{4\Omega}{\lambda^2}
\right] \label{eq:mass_small_L_0}\\
&\simeq& \frac{2}{3} g_5^2  \,\Omega \left(2v^2 + \bar v^2\right) L^2\,,\qquad(\Omega L \ll 1)\,,
\label{eq:mass_small_L}
\ee
where we have used
$\frac{1}{g_5^2} =  \frac{4a\bar v^2}{\Omega}$ as is obtained in Eq.~(\ref{eq:4dim_g}),
$\big<\gamma^{(0)}\big|\gamma^{(0)}\big> = \frac{4\bar v^2}{\Omega}$,
and $\tilde N$ is given in Eq.~(\ref{eq:tN}).
This approximation is compared with numerically obtained masses in Fig.~\ref{fig:mass_off_diagonal}.
They nicely match for $\Omega L \ll 1$.
This behavior resembles the standard Higgs mechanism 
that mass of the vector boson 
is a product of a gauge coupling and a scalar vacuum expectation value (VEV).
The mass formula (\ref{eq:mass_small_L}) tells that the effective VEV is
\be
v_{\rm eff} = \sqrt{\frac{2}{3}  \,\Omega \left(2v^2 + \bar v^2\right)}\, L\,,
\ee
so that the mass is given by 
$\tilde \mu_0(L)^2= g_5 v_{\rm eff}$.

Note that peculiar phenomena in the geometric
Higgs mechanism in our specific model appear in the opposite limit $L\Omega \gg 1$.
Indeed, 
as $L$ being increased, the
bottom of the well is lifted, and high potential barrier appears at the center of the domain walls. 
Height of the potential is exponentially large as
\be
{\cal V}(0) = \frac{\cosh\Omega L-1}{a}+2 \Omega^2 
-3\Omega^2\sech^2\frac{L\Omega}{2} \quad \to\quad
\frac{e^{L\Omega}}{2a}\qquad  \text{as}\quad L \to \infty\,. 
\ee
Therefore, for large separations $\Omega L \gg 1$, as shown in Fig.~\ref{fig:sch_pot_off},
the massive vector bosons are located not between but on the domain walls, and
their masses asymptotically approach to threshold mass  
$\sqrt{\Omega^2+1/a}$ which can be read from 
${\cal V}(\pm\infty) = \Omega^2 + \frac{1-\sech\Omega L}{a} 
\sim \Omega^2 + 1/a$. Thus, when the wall separation is large, the mass of vector boson becomes universal which is about $\sqrt{\Omega^2 + 1/a}$
irrespective of $L$, which is of order $M_5$.

Finally, we have to solve the coupled equations (\ref{eq:leom_f}) and (\ref{eq:leom_sc}) for the divergence part $\tilde b_{\rm d}^\mu$
and $\bar{\bm t}$.  Let us introduce 
\be
\p^\mu B &\equiv&  \sigma_+\,\p^\mu \frac{\p^\nu}{\p^2}  \tilde b_\nu = \sigma_+ \tilde b_{\rm d}^\mu \,,\\
\bar{\bm t}' &\equiv& \bar{\bm t} - \frac{i}{\tilde N} \frac{B}{\sigma_+} \otimes \tilde{\mathbbm{u}}^{(0)}\,.
\ee
There is redundancy that any function of $y$ can be added to $B$.
Now Eqs.~(\ref{eq:leom_f}) and (\ref{eq:leom_sc}) are rewritten as
\be
\p^\mu\left[ a \sigma_+
\tilde Q^\dagger \tilde Q B
+ \frac{i}{\tilde N} \tilde{\mathbbm{u}}^{(0)t} \, \bar{\bm t}' \right] = 0\,,\\
\left(-\p_y^2 + \tilde V\right) \left(\bar{\bm t}' + \frac{i}{\tilde N}\frac{B}{\sigma_+}\otimes \tilde{\mathbbm{u}}^{(0)}\right) 
= \p^2 \bar{\bm t}'\,,
\label{eq:off_diag_scalar_2}
\ee
where $\tilde Q$ and $\tilde Q^\dagger$ are defined in Eq.~(\ref{eq:QQ}).
The former equation can formally be solved as
\be
B =\left(\tilde Q^\dagger \tilde Q\right)^{-1} 
\left(\frac{-i}{a\tilde N\sigma_+}\tilde{\mathbbm{u}}^{(0)t}\,\bar{\bm t}' + \Lambda(y)\right)\,,
\label{eq:B}
\ee
where $\Lambda(y)$ is an arbitrary function of $y$ which we may set 0 by absorbing in the redundancy of $B$.
Plugging this into Eq.~(\ref{eq:off_diag_scalar_2}), one can eliminate $B$ and we are left with
\be
\p^2 \bar{\bm t}' = \left(-\p_y^2 + \tilde V\right)\left\{\bar{\bm t}'
+ \frac{i}{\tilde N}\frac{1}{\sigma_+}
\left[\left(\tilde Q^\dagger \tilde Q \right)^{-1} 
\left(\frac{-i}{a\tilde N\sigma_+}\tilde{\mathbbm{u}}^{(0)t}\,\bar{\bm t}' \right)\right]\otimes \tilde{\mathbbm{u}}^{(0)}
\right\}\,.
\label{eq:bar_t'}
\ee
This formally determines the mass spectrum for $\bar{\bm t}'$ although it is not easily solved analytically.
Once we do this, $B$ is determined from Eq.~(\ref{eq:B}). Thus, as usual, the divergence part $\tilde b_{\rm d}^\mu$ does not
have independent physical degrees of freedom, so that Eq.~(\ref{eq:leom_f}) should be regarded as the constraint reducing 
the four polarization degrees of freedom in a massive vector field by one.

Instead of trying to solve Eq.~(\ref{eq:bar_t'}),  let us understand the spectrum defined by
seeing Eq.~(\ref{eq:bar_t'}) from a different viewpoint. Let us first recall that the second term on the right-hand side of 
Eq.~(\ref{eq:leom_sc}) or (\ref{eq:bar_t'}) reflects the fact that we gauged $SU(5)$ symmetry of the corresponding mode
equation (\ref{eq:leom_t_scalar2}) in the scalar model considered in Sec.~\ref{sec:dwzm}.
As was mentioned at the end of Sec.~\ref{sec:dwzm}, when $SU(5)$ is global symmetry,
the four-dimensional effective fields in $\tilde{\bm{t}}$ (or $\bar{\bm{t}}$) are
all massless NGs at the infinite wall separation, and they, except for the genuine NGs, are lifted and obtain non-zero mass
of order the inverse separation $\sim 1/L$ when the walls are separated by $L$. This is similar to the standard compactification
of the fifth direction to $S^1/Z_2$ with radius $R$. In our theory, the extra dimension is infinitely large and our compactification is
a posteriori done by the domain walls with the compactification size $L$. 
On the other hand, in the $S^1/Z_2$ model the extra dimension is compact a priori.  
For simplicity, let us compare the simplest models, the $U(1)$ Goldstone (global) model and Abelian-Higgs (gauge) model
in five dimensions compactified by $S^1/Z_2$ orbifolding,
\be
{\cal L}_{\rm global} &=& |D_M \phi|^2 - \frac{\lambda^2}{4}\left(|\phi|^2 - v^2\right)^2\,,\\
{\cal L}_{\rm local} &=& -\frac{1}{4} F_{MN}F^{MN} + |D_M \phi|^2 - \frac{\lambda^2}{4}\left(|\phi|^2 - v^2\right)^2\,.
\label{eq:L_AH}
\ee
If the compactification radius $R$ is infinite, the spectrum is $0$ and $\sqrt 2 \lambda v$ in the global
model, and $ev$ and $\sqrt 2 \lambda v$ in the gauged model. These levels are infinitely degenerate in the four-dimensional sense.
They are split when the compactification radius $R$ is finite. The spectra in Feynman gauge are split into
$0$, $k/R$, $\sqrt 2 \lambda v$ and $\sqrt{2\lambda^2 v^2 + (k/R)^2}$ in the former model while
$ev$, $\sqrt{e^2v^2 + (k/R)^2}$, $\sqrt 2 \lambda v$ and $\sqrt{2\lambda^2 v^2 + (k/R)^2}$ in the latter model, where $k$ is an integer
for the Kaluza-Klein (KK) tower, see appendix \ref{app:KKAH} for  details. For our purpose of understanding Eq.~(\ref{eq:bar_t'}), we 
emphasize the fact that the KK tower $\{0,k/R\}$ of the NG modes is shifted to $\{ev,\sqrt{e^2v^2 + (k/R)^2}\}$ by gauging $U(1)$ symmetry.
Namely, the light modes of order $1/R$ in the global model acquire heavy masses of order $ev$ (the five-dimensional mass scale) by
gauging. This should happen also in our model because difference between our model and the orbifold model is how to compactify the
extra dimension.
Remember again that the spectrum of $\tilde{\bm{t}}$ in our scalar model treated in Sec.~\ref{sec:dwzm} consists of
the massless NGs and the massive modes of order $1/L$. When we gauge $SU(5)$ symmetry, the massless NGs are eaten 
by the geometric Higgs mechanism to give the mass $\tilde \mu_0(L)$ 
(it is of order the fundamental mass scale in five dimensions because of $\tilde \mu_0(L) \to \sqrt{\Omega^2 + 1/a}$ for $\Omega L \gg 1$) 
to the off-diagonal gauge fields. Therefore, the massive modes of the order $1/L$ in the scalar model acquire heavy mass of the order
$\sqrt{(1/L)^2 + \tilde \mu_0^2}$ by gauging. In conclusion, there are no modes below $\tilde \mu_0$ in the $\tilde{\bm{t}}$ channel, and
therefore we do not worry about phenomenologically undesired light modes from the off-diagonal elements.

\subsection{Field theoretical D-branes}

It is worthwhile pointing out that the number of coincident walls corresponds to the rank of the gauge group preserved by
the domain wall configurations. When $k$ domain walls coincide, massless $SU(k)$ gauge fields are localized there.
This is quite similar to D-branes in superstring theory.
Indeed, in addition to the $SU(k)$ gauge fields, two massless scalar fields from $T$ and $S$ 
in the adjoint representation of $SU(k)$ are localized in our model. This resembles bosonic component of ${\cal N}=4$ $SU(k)$ vector multiplets
appearing at $k$ coincident D3-branes, though we have no fermions and additional four adjoint scalar fields are needed.
Furthermore, the mass formula given in Eq.~(\ref{eq:mass_small_L}) 
for $\Omega L \ll 1$ tells that the mass of lightest vector bosons
is proportional to the wall separation $L$. This is similar to the fact that massive vector boson on the separated D-branes is proportional
to D-brane separation because its origin is F-strings stretching 
between separated D-branes.
Thus, the domain walls in our model with the geometric Higgs mechanism strongly 
resembles similar mechanism in D-brane physics.

\section{Non-linear effective Lagrangian for zero modes}\label{sc:effective_action2}

In this section, we derive a low-energy effective 
Lagrangian in the so-called moduli approximation \cite{Manton}, 
where the moduli parameters of the background solution are 
promoted to slowly varying fields. 
In other words,  we promote the $N\times N$ matrix of parameters $Y$ in the general solution 
Eq.~\refer{eq:gensol3} to four-dimensional fields $Y \to \Omega^{-1} \mathbf{Y}(x)$ which transform as an adjoint under $SU(N)$. 
Note that we set $\mathbf{Y}$ dimensionless for maintaining simplicity in the following calculations. 
As a result, the five-dimensional scalar fields become 
functions of the effective four-dimensional moduli fields:
\begin{align}
T(x,y) & = v \tanh\bigl(\Omega y\mathbf{1}_N-\mathbf{Y}(x)\bigr)\,, 
\label{eq:moduli-approx-T}
\\
S(x,y) & = \bar v\, \sech\bigl(\Omega y\mathbf{1}_N-\mathbf{Y}(x)\bigr)\,.
\label{eq:moduli-approx-S}
\end{align}
The goal of this section is to describe effective four-dimensional dynamics of $\mathbf{Y}(x)$. 
We present  the metric of the moduli space, which gives 
full non-linear interaction of moduli fields in a closed form. 
We limit ourselves to terms with at most two derivatives, although 
we can compute higher derivative corrections with increasing 
complexity \cite{Eto:2012qda,Eto:2014gya}. 

\subsection{Effective Lagrangian in the $k_1$-$k_2$ split background}

To illustrate our approach, let us first present the effective 
Lagrangian in a fixed background with $k_1$-$k_2$ split 
configuration of walls. 
Furthermore, we will first restrict ourselves only 
to the leading order effects in moduli fields to keep the discussion simple.
However, in the next subsection, we will present a closed formula for the effective Lagrangian which captures all non-linear interactions of moduli and works in arbitrary background.

To pick up the $k_1$-$k_2$ background, we assume that $\mathbf{Y}(x)$ is decomposed as
\begin{align}\label{eq:32split}
\mathbf{Y}(x) & = \begin{pmatrix}
{\mathcal Y}_{k_1} \mathbf{1}_{k_1} & \mathbf{0}_{k_1\times k_2} \\
\mathbf{0}_{k_2\times k_1} & {\mathcal Y}_{k_2} \mathbf{1}_{k_2}
\end{pmatrix}+
\begin{pmatrix}
\mathbf{Y}_{k_1}(x) & \mathbf{0}_{k_1\times k_2} \\
\mathbf{0}_{k_2\times k_1} & \mathbf{Y}_{k_2}(x)
\end{pmatrix}.
\end{align}
Here, parameters ${\mathcal Y}_{k_1}$ and ${\mathcal Y}_{k_2}$ 
are positions of the $k_1$ and the $k_2$ of walls respectively, 
while $\mathbf{Y}_{k_1}(x)$ and $\mathbf{Y}_{k_2}(x)$ are fields 
transforming under $S[U(k_1)\times U(k_2)]$ gauge group. 
From the point of view of the effective theory we can think 
of the first part of this decomposition as a 
`vacuum expectation value' of $\mathbf{Y}(x)$, while the second 
part represents the fluctuations. 
The vacuum expectation value ${\mathcal Y}_{k_1} \not = 
{\mathcal Y}_{k_2}$ is what determines the symmetry breaking pattern.
In this sense, the geometric Higgs mechanism of the five-dimensional 
theory is similar to an ordinary Higgs mechanism 
in four-dimensional theory with $\mathbf{Y}(x)$ playing a role 
of an adjoint Higgs field. 

Notice that off-diagonal components in the second part of the decomposition \refer{eq:32split} are set to zero. The physical reason is  the Higgs mechanism. More precisely, we can always absorb these fields into a definition of the corresponding off-diagonal components of gauge fields by an appropriate gauge transformation. In other worlds, in the decomposition \refer{eq:32split} we are assuming the so-called unitary gauge where only physical fields appear.

Next, let us consider the gauge fields. We have established that the wave-function of massless gauge fields is flat, hence we can just replace $A_\mu(x,y) \to A_\mu(x)$. However, we also need to decompose the $SU(N)$ gauge fields into the $S[U(k_1)\times U(k_2)]$ fields:
\begin{align}\label{eq:fielddec}
A_\mu & = \begin{pmatrix}
a_{k_1\mu} &  \mathbf{0}_{k_1\times k_2} \\ 
 \mathbf{0}_{k_2\times k_1} &  a_{k_2\mu}
\end{pmatrix}+ a_{1\mu}T_1\,, \\
G^{\mu\nu} & = \begin{pmatrix}
f_{k_1}^{\mu\nu}  & \mathbf{0}_{k_1\times k_2} \\
 \mathbf{0}_{k_2\times k_1}  & f_{k_2}^{\mu\nu}
\end{pmatrix} + f_1^{\mu\nu}T_1\,,
\end{align}
where $a_{i\mu}$ and $f_{i\mu\nu}$ ($i=1,k_1,k_2$) are massless gauge bosons and field strengths of the respective $SU(i)$ gauge groups,
while 
\begin{eqnarray}
T_1= \sqrt{\frac{k_1k_2}{2N}}\,\mbox{diag}\left(\frac{1}{k_1},\cdots,\frac{1}{k_1},-\frac{1}{k_2},\cdots,-\frac{1}{k_2}\right)
\end{eqnarray}
 is the hypercharge generator. We do not include the off-diagonal components in the decomposition \refer{eq:fielddec} as these represent massive vector bosons and hence, in the spirit of the low-energy limit, we ignore them.

The effective Lagrangian is obtained by inserting all the above decompositions into the full Lagrangian and integrating it over the $y$-axis. 
Neglecting higher than second powers of moduli fields we get
\begin{align}
{\mathcal L}_{\footnotesize \mbox{eff}} & = - E -\frac{1}{2g_{k_1}^2}\Tr\bigl[f_{k_1\mu\nu}f_{k_1}^{\mu\nu}\bigr]
 -\frac{1}{2g_{k_2}^2}\Tr\bigl[f_{k_2\mu\nu}f_{k_2}^{\mu\nu}\bigr]-\frac{1}{4g_1^2}f_{1\mu\nu}f_1^{\mu\nu} \nonumber \\
 & + 
 \frac{T_W}{2}\Tr\bigl[D_\mu \mathbf{Y}_{k_1}D^{\mu}\mathbf{Y}_{k_1}+ D_\mu \mathbf{Y}_{k_2}D^{\mu}\mathbf{Y}_{k_2}\bigr] 
 +O\bigl(\mathbf{Y}_{k_1}^3,\mathbf{Y}_{k_2}^3\bigr)\,.
  \label{eq:totallagr32}
\end{align}
The contributions without four-dimensional derivatives 
sum up to the first term $E = N T_W$, which is a constant 
(topological charge) equal to the total tension of walls given in Eq.~(\ref{eq:wall_tension}) and 
has no effect on dynamics. Further, we have the gauge couplings
\begin{gather}
\frac{1}{g_{k_1}^2} =  \frac{1}{g_{k_2}^2} = \frac{1}{g_1^2} = \frac{4a \bar v^2}{\Omega}\,. \label{eq:gaugerel}
\end{gather}
This corresponds to those in Eq.~(\ref{eq:4dim_g}), though here we are considering generic $N$ and $k_1$ ($k_2 = N-k_1$).

The factor standing in front of kinetic terms of moduli fields $\mathbf{Y}_{k_1}$ and $\mathbf{Y}_{k_2}$ is equal to $\tfrac{1}{2}T_W = \tfrac{1}{2}E/N$ or, in other words, a half of the tension of a single domain wall. This is to be expected, since the same is true for translational pseudo-NG zero mode of any domain wall. Indeed, both  $\mathbf{Y}_{k_1}$ and $\mathbf{Y}_{k_2}$ contain translational moduli of the respective $k_1$- and $k_2$-plets of walls.

The effective Lagrangian \refer{eq:totallagr32}, while a correct 
four-dimensional description of the moduli dynamics in the $k_1$-$k_2$ 
split background, has an unsatisfactory feature. 
It breaks down in the limit ${\mathcal Y}_{k_1} \to {\mathcal Y}_{k_2}$, 
where the $SU(N)$ gauge invariance is restored. 
Indeed, at the coincident point we have more massless fields 
than those appearing in Eq.~\refer{eq:totallagr32}, namely 
off-diagonal components of gauge fields and moduli fields. 
It would be more appropriate to have an effective theory which can continuously transit from one breaking pattern to another and simultaneously keep track of all fields. Fortunately, this can be done by fully utilizing the moduli approximation as we will see below.   

\subsection{The Extended Effective Lagrangian in the arbitrary background}

Let us adopt the same ansatz for scalar fields as in Eqs.~\refer{eq:moduli-approx-T}-\refer{eq:moduli-approx-S}. This time, however, we will not assume any particular background and leave the $N\times N$ adjoint moduli fields $\mathbf{Y}(x)$ completely arbitrary.
The gauge fields $A_\mu$ are given by their zero modes 
(in axial gauge), which happen to be independent of moduli fields. 
\begin{align}
A_\mu(x,y) & = A_\mu(x)\,, 
\label{eq:moduli-approx-amu}
\\
A_y(x,y) & = 0\,. 
\label{eq:moduli-approx-ay}
\end{align}
Since we work in arbitrary background, there is no apriori distinction between unbroken and broken generators. 
Hence, the formula \refer{eq:moduli-approx-amu} keeps track of all gauge bosons, contrary to our discussion in the previous subsection, 
where we discarded the off-diagonal massive fields.
In what follows, we assume that all $N^2-1$ components of $A_\mu$ have a flat wave-function along the $y$ axis. 
This is evidently an approximation, which is forced on us by the fact that we were not able to derive a closed analytic formula 
for the wave-function of massive vector fields. 
Indeed, we learned in Sec.~\ref{sec:geomhiggs} that we can determine $\gamma^{(n)}(y)$  only numerically.
If we had such an analytic formula, 
it would be possible to improve Eq.~\refer{eq:moduli-approx-amu} to accommodate for moduli-dependent effects. 
 
The effective Lagrangian is obtained by plugging the ansatz 
into the five-dimensional Lagrangian (\ref{eq:lagrangian}) and 
integrating it over the $y$-axis. 
 In carrying out the calculations we employ identities which are gathered in the Appendix~\ref{app:A}. 
The result reads in the following closed form:
\begin{align}\label{eq:totallagr}
{\mathcal L}_{\footnotesize \mbox{eff}} & = 
-E -\frac{1}{2g_5^2}\Tr\bigl(G_{\mu\nu}G^{\mu\nu}\bigr) +
\Tr\Bigl\{g\bigl({\mathcal L}_{\mathbf{Y}}\bigr)
\bigl[D_\mu\mathbf{Y}\bigr]\,D^{\mu}\mathbf{Y}\Bigr\}\,,
\end{align}
where ${\mathcal L}_{\mathbf{Y}}[\cdot] \equiv [\mathbf{Y},\cdot]$ is a Lie derivative and where
\begin{align}\label{eq:gfun}
g(\beta) & = \frac{4}{\Omega\, \beta^2}\left(v^2 \beta\, \mbox{coth}(\beta)-\bar v^2\beta\, \mbox{cosh}(\beta)-\frac{\Omega^2}{\lambda^2}\right)\,.
\end{align}
This is the main result of this section. The effective Lagrangian  \refer{eq:totallagr} captures full non-linear interaction of moduli fields to all orders and it can be adopted to any background. 
For example, we can describe continuous transition from the fully coincident configuration to $k_1$-$k_2$ split configuration by decomposing moduli fields as in Eq.~\refer{eq:32split}. In the limit ${\mathcal Y}_{k_1} \to {\mathcal Y}_{k_2}$ we have unbroken $SU(N)$ gauge symmetry and all gauge fields are massless. Once we depart from this point, off-diagonal components of gauge fields, which are denoted by $k_1\times k_2$ complex matrix $b_\mu$, become massive. 

In order to compare this with the results in Sec.~\ref{sec:geomhiggs}, let us consider $N=5$ and $(k_1,k_2) = (3,2)$.
In the effective Lagrangian, their mass term arise as the leading term in the expansion in terms of moduli fields of 
\begin{multline}
\frac{1}{2g_5^2}\int_{-\infty}^{\infty}dy\, \Tr\Bigl[\bigl[A_\mu, T\bigr]\bigl[A^\mu, T\bigr]+\bigl[A_\mu, S\bigr]\bigl[A^\mu, S\bigr]\Bigr] = \\ 
\frac{1}{g_5^2}
\left\{4v^2 L \tanh\frac{L\Omega}{2}-\frac{4\Omega}{\lambda^2}\left(1-\frac{L\Omega}{\sinh L\Omega}\right)\right\}\Tr\bigl[b^\mu b_\mu^{\dagger}\bigr]+\ldots 
\end{multline}
where $L \equiv \frac{{\mathcal Y}_3-{\mathcal Y}_2}{\Omega}$ and 
where ellipses indicates a higher order corrections describing interaction of $b_\mu$ with moduli fields. 
Now, we can read the mass of massive gauge boson $b_\mu$,
\begin{equation}
\tilde \mu_0^{\rm (eff)}(L)^2 = \frac{1}{g_5^2}\left[4v^2 L \tanh\frac{L\Omega}{2}-\frac{4\Omega}{\lambda^2}\left(1-\frac{L\Omega}{\sinh L\Omega}\right)\right]\,.
\end{equation}
Note that this precisely coincides with $\tilde \mu_0(L)$ at $\Omega L \ll 1$ given in Eq.~(\ref{eq:mass_small_L}). 
Thus, the utility of the effective Lagrangian \refer{eq:totallagr} is maximal when walls are close to each other. 
For $\Omega L \gg 1$, 
this mass, of course, differs from the true mass of $b_\mu$ due to the fact that our assumption, the flat wave-functions $\gamma^{(0)}$,
breaks down. 

Note that, as we have seen in Eq.~(\ref{eq:totallagr32}), 
usually the moduli approximation can deal with only the massless fields and can describe their dynamics at energy scale much
below that of the original theory, say $M_5$ in this work.
We should emphasize that 
the extended effective Lagrangian (\ref{eq:totallagr}) can 
describe dynamics of not only the massless fields but also 
the massive fields. 
This is quite natural that the mass of gauge boson is 
proportional to the wall separation $L$, hence it can be 
arbitrarily small. Therefore, the moduli approximation should detect their presence as long as $\Omega L \ll 1$,
and indeed (\ref{eq:totallagr}) can do it. Thus, we have proven with the extended effective Lagrangian (\ref{eq:totallagr})  that the geometric
Higgs mechanism occurs at the level of the low energy effective theory.

The third term of the effective Lagrangian \refer{eq:totallagr} contains kinetic terms for moduli fields, which exhibit non-trivial self-interaction. This reflects the fact that the moduli space is curved and that the zero modes move along the geodetics \cite{Manton}.
The metric of the moduli space can in fact be easily calculated. Let us decompose $\mathbf{Y}$ into generators of $U(N)$ as $\mathbf{Y} = Y_I T_I = Y_0\mathbf{1}_5+Y_{a}T_a$, where 
\begin{equation}
\comm{T_a}{T_a}=if_{abc}T_{c}\,, \hspace{5mm} \Tr\bigl[T_{a}T_{b}\bigr] = \frac{1}{2}\delta_{ab}\,.
\end{equation}
The functions $Y_0$ and $Y_a$, $a=1,\ldots N^2-1$ can be treated as independent zero modes.
The metric on the moduli space is then given as overlap between them:
\begin{align}
g_{IJ} & = 2v^2\int_{-\infty}^\infty dy\, \Tr\Bigl[\partial_I\tanh\bigl(\Omega y\mathbf{1}_N-\mathbf{Y}\bigr)\partial_J\tanh\bigl(\Omega y\mathbf{1}_N-\mathbf{Y}\bigr)\Bigr] \nonumber \\ & +2\bar v^2 \int_{-\infty}^\infty dy\, \Tr\Bigl[\partial_I\,\sech\bigl(\Omega y\mathbf{1}_N-\mathbf{Y}\bigr)\partial_J\,\sech\bigl(\Omega y\mathbf{1}_N-\mathbf{Y}\bigr)\Bigr] \nonumber \\
& = 2 \Tr\bigl[g\bigl({\mathcal L}_{\mathbf{Y}}\bigr)[T_I]\,T_J\bigr]\,,
\end{align}
where $\partial_I Y_J = \delta_{IJ}$ and where $g({\mathcal L}_{\mathbf{Y}})$ is same as in Eq.~\refer{eq:gfun}. Using the identity \refer{eq:gident} from the Appendix, we can write down the metric explicitly as
\begin{gather}
g_{00} = NT_W \,, \hspace{5mm} g_{0a} =0\,, \\
g_{ab} = \frac{4}{\Omega}\Bigl(\bigl(v^2-\bar v^2\bigr)\mathbf{\tilde Y}^{-2}-v^2\mathbf{\tilde Y}^{-1}\mbox{tan}^{-1}\bigl(\mathbf{\tilde Y}\bigr)+\bar v^2\mathbf{\tilde Y}^{-1}\mbox{sin}^{-1}\bigl(\mathbf{\tilde Y}\bigr)\Bigr)_{ab}\,,
\end{gather}
where $\mathbf{\tilde Y}$ is a $N^2-1\times N^2-1$ matrix with elements $\mathbf{\tilde Y}_{ab} = f_{abc}Y_c$. With the above metric we can rewrite the third term of ${\mathcal L}_{\footnotesize \mbox{eff}}$ into the compact form
\begin{equation}
\frac{1}{2}g_{IJ}\bigl(D_\mu Y\bigr)_I\bigl(D^\mu Y\bigr)_J\,,
\end{equation}
where
\begin{equation}
\bigl(D_\mu Y\bigr)_I = \left\{\begin{array}{cc}
I = 0 & \partial_\mu Y_0 \\
I = a & \partial_\mu Y_a +if_{abc}A_bY_c
\end{array}\right.
\end{equation}

Although we only consider terms up to two derivatives, it is 
believed that effective dynamics of moduli fields of domain 
walls can be also captured by Nambu-Goto type action or, more 
generally, as a function of Nambu-Goto action \cite{Abraham:1992vb,Gauntlett:2000de,Eto:2015vsa,Hashimoto}. 
To our best knowledge, there seems to be no solid consensus about 
how to extend Nambu-Goto action to accommodate non-Abelian 
symmetry or multi-wall configurations. 
The results of this section could potentially be relevant for 
these efforts, especially if they are supplemented by 
four-derivative corrections.

\section{Conclusions and Discussions
}\label{sec:discussion}

In this paper we presented a (4+1)-dimensional model which 
gives a framework of dynamical realization of the brane world model by domain walls, 
incorporating 
the two core ideas: 
the semi-classical localization mechanism for gauge fields 
and geometric Higgs mechanism using a multi-domain wall 
background.
Since the domain walls interpolate multiple vacua which preserve different subgroups of $SU(N)$,
multiple Higgs mechanisms occur locally at the same time. As the domain walls are smooth and continuous
solutions of the field equations, the local Higgs mechanisms should be smoothly connected. This is the geometric Higgs
mechanism which we investigated in detail in this work. The off-diagonal vector bosons get nonzero masses by
eating the non-Abelian clouds which are localized moduli of the multiple domain wall solutions.
In this work, we investigated this phenomenon and evaluated the mass of the vector boson 
by analyzing the mass spectra from the (4+1)-dimensional viewpoint in
Sec.~\ref{sec:threetwo}. We also confirmed the geometric Higgs mechanism from the perspective of low energy 
effective theory on the domain walls in Sec.~\ref{sc:effective_action2}. 
Through the analysis, we extended conventional moduli approximation \cite{Manton} to the theory which naturally include 
not only massless modes but also massive modes via the geometric 
Higgs mechanism, provided the masses are much less than the mass gap 
of the $(4+1)$-dimensional theory.

Although we have not dealt with grand unification theories (GUT) at all in this paper, natural and important application of
the geometric Higgs mechanism is, doubtless, to realize GUT dynamically on the domain walls. We will investigate
it separately in the subsequent work \cite{2ndpaper}.

As to the other side of our result, we pointed out deep similarity between our domain walls and D-branes in superstring theories.
This similarity goes beyond the often-cited connections between the field theoretical solitons and D-branes.
The similarities are three-folds. First, the number $k$ of coincident walls corresponds to the rank of the special unitary gauge group $SU(k)$.
Second, the mass of massive gauge boson is proportional to the separation of the walls at least if the separation is sufficiently small.
Third, the field content appearing on the $k$ coincident domain walls are a subset of $SU(k)$ vector multiplet of ${\cal N}=4$ supersymmetric Yang-Mills
theory that is the low energy effective theory of $k$ coincident D3-branes.
The main reason behind these similarities is the localization of non-Abelian gauge fields which is  caused by the confinement in the bulk 
realized semi-classically via the field-dependent gauge kinetic term \cite{Ohta}.

In this paper, we have not taken SUSY as our guiding principle. 
However, it may be advantageous to consider five-dimensional SUSY 
with eight supercharges as a master theory. 
The immediate benefit of implementing SUSY is that a 
gauge kinetic term as a function of scalar field
occurs naturally via the prepotential \cite{Ohta}. 
Further, domain walls are often realized as 1/2 BPS objects, 
which spontaneously break half of the supercharges. 
A combination of wall and anti-wall in the background then 
breaks SUSY entirely \cite{Maru:2001gf,Maru:2000sx}.

\begin{figure}
\begin{center}
\includegraphics[width = 0.5\textwidth]{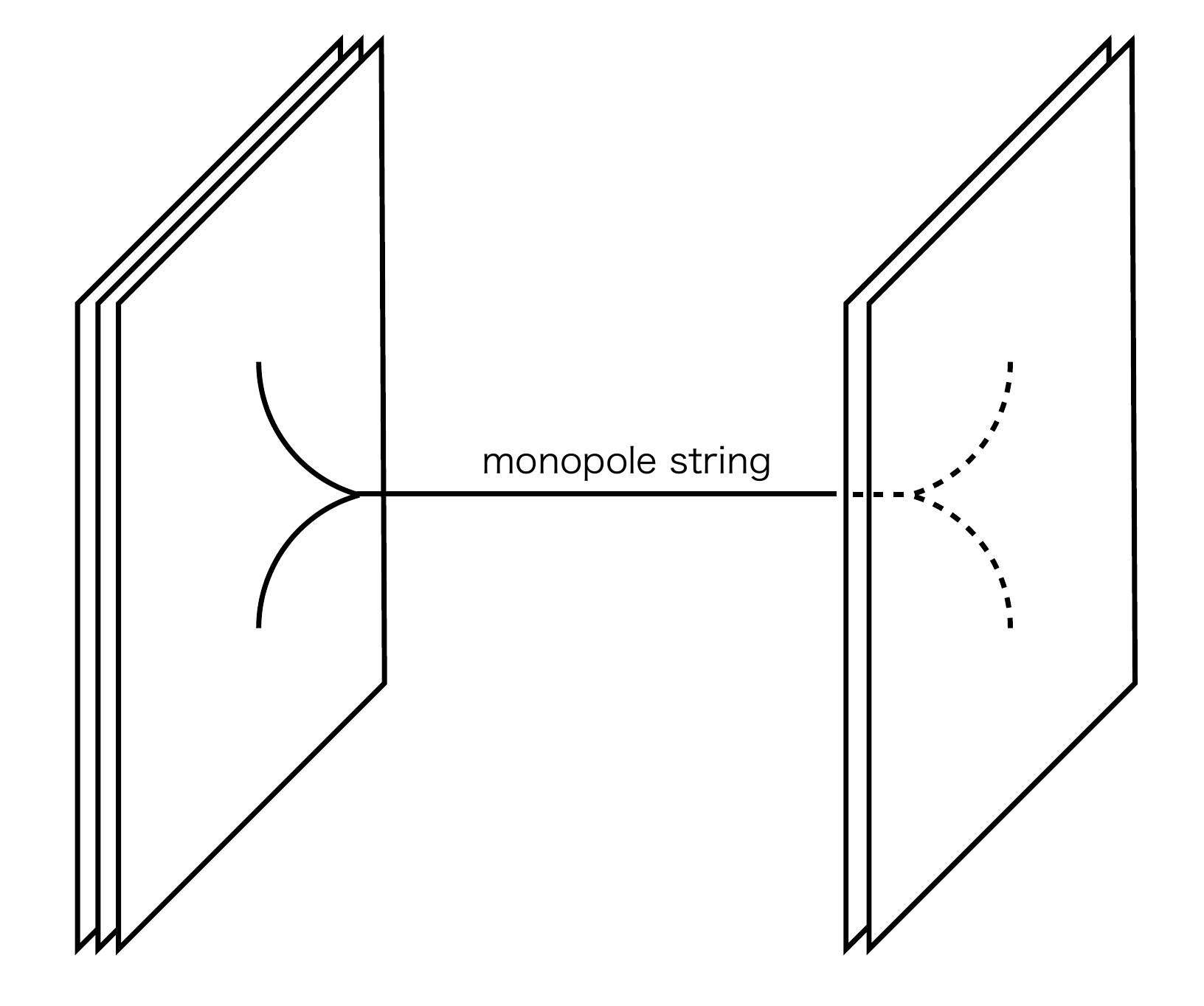}
\caption{\small A schematic depiction of the magnetic monopole 
from the 5-dimensional point of view as a string stretched 
between separated domain walls.}
\label{fig:monopole}
\end{center}
\end{figure}

Another feature common to most non-Abelian gauge theories is magnetic 
monopoles, which originate from the breaking of the 
semi-simple gauge group to a subgroup with a $U(1)$ factor. 
From a phenomenological point of view, monopoles are 
important in cosmology.
In standard Yang-Mills theory in 3+1 dimensions, the magnetic monopoles are 
't Hooft-Polyakov type point-like solitons \cite{'tHooft, Polyakov}. 
However, in our model monopoles arise as string-like 
objects, stretched between separated domain walls, as depicted 
in Fig.~\ref{fig:monopole}. 
The reason is that the asymptotic vacua outside domain walls 
are $SU(N)$ preserving, while only between the walls the 
symmetry is broken down to some subgroup, allowing for 
a non-trivial topology.
Let us also remark, that this type of configuration has a direct 
analog in D-strings \cite{Diaconescu}. 
Detailed study of these observations is a subject of a 
forthcoming work.

 Lastly, in this work we have not discussed gravity for simplicity. 
However, an interesting direction for future study may be to 
consider Randall-Sundrum-like theory \cite{RS, RS2,Eto:2002ns, Arai:2002ph, Eto:2003ut, Eto:2003xq,Eto:2003bn,Eto:2004yk} with multiple 
branes and investigate the issues of cosmological constant and 
the hierarchy problem there. 
Another and a perhaps more natural option is to employ 
position-dependent gravitational constant as a means to 
localize massless gravitons on multiple domain walls, thus 
creating the background dynamically. 
In this setting, it would be interesting to investigate spectra 
of graviphotons and fluctuations of domain walls as they are 
natural candidates for dark matter (for details of massive 
vector-like dark matter coming from brane oscillations see 
\cite{Clark1, Clark2}). We plan to elaborate on this in the near future.

\acknowledgements

F.\ B.\ thanks M.\ Nitta for useful discussions and comments. 
F.\ B.\ was an international research fellow of the Japan Society 
for the Promotion of Science, and was supported by Grant-in-Aid 
for JSPS Fellows, Grant Number 26004750. 
This work is also supported in part 
by the Ministry of Education,
Culture, Sports, Science (MEXT)-Supported Program for the Strategic
Research Foundation at Private Universities ``Topological Science''
(Grant No.~S1511006), 
by the Japan Society for the 
Promotion of Science (JSPS) 
Grant-in-Aid for Scientific Research
(KAKENHI) Grant Numbers 25400280 (M.A.),  
26800119 and 16H03984 (M.\ E.), and 
25400241 (N.\ S.), and by the Albert Einstein Centre for Gravitation and Astrophysics financed by the Czech Science Agency Grant No. 14-37086G (F.\ B.).

\appendix

\renewcommand{\theequation}{A.\arabic{equation}}
\section{Kaluza-Klein expansion of Abelian-Higgs model}
\label{app:KKAH}

Let us consider Abelian-Higgs model in 5 dimensions
\be
{\cal L}_5 = - \frac{1}{4}F_{MN}F^{MN} + |D_M \phi|^2 - \frac{\lambda^2}{4}\left(|\phi|^2 - v^2\right)^2.
\ee
We compactify the fifth direction to $S^1/Z_2$ with radius $R$, and we impose periodicity to all fields as $f(y+2\pi R) = f(y)$.
The KK expansions for the scalar and vector fields are given by
\be
\phi(x,y) &=& v + \frac{1}{\sqrt{2\pi R}} \left(\eta^{(0)}(x) + i \sigma^{(0)}(x)\right) 
+ \frac{1}{\sqrt{\pi R}} \sum_{k=1}^\infty  \left(\eta^{(k)}(x) + i \sigma^{(k)}(x)\right)\cos \frac{k}{R}y,\\
A_\mu(x,y) &=& \frac{1}{\sqrt{2\pi R}} a_\mu^{(0)}(x) 
+ \frac{1}{\sqrt{\pi R}} \sum_{k=1}^\infty  a_\mu^{(k)}(x) \cos \frac{k}{R}y,\\
A_y(x,y) &=&  \frac{1}{\sqrt{\pi R}} \sum_{k=1}^\infty  a_y^{(k)}(x) \sin \frac{k}{R}y ,
\ee
where we imposed that $\phi$ and $A_\mu$ are even whereas $A_5$ is odd for the orbifold parity. 
Let us write down the quadratic Lagrangian by inserting these KK expansions into ${\cal L}_5$ and integrating it by $y$.
Then we get
\be
-\frac{1}{4}\int^{2\pi R}_0 dy~F_{\mu\nu}F^{\mu\nu} = -\frac{1}{4} f^{(0)}_{\mu\nu}f^{(0)\mu\nu}
- \frac{1}{4} \sum_{k=1}^\infty 
f^{(k)}_{\mu\nu}f^{(k)\mu\nu},
\ee
with $f^{(k)}_{\mu\nu} = \p_\mu a_\nu^{(k)} - \p_\nu a_\mu^{(k)}$. We also have
\be
- \frac{1}{2}\int^{2\pi R}_0dy~F_{\mu y} F^{\mu y} = \frac{1}{2} \sum_{k=1}^\infty 
\left(\p_\mu a_y^{(k)}+ \frac{k}{R} a_\mu^{(k)}\right)^2
\ee
where we have used 
\be
F_{\mu y} =
\frac{1}{\sqrt{\pi R}} \sum_{k=1}^\infty  \p_\mu a_y^{(k)} \sin \frac{k}{R}y
+ 
\frac{1}{\sqrt{\pi R}} \sum_{k=1}^\infty \frac{k}{R} a_\mu^{(k)}\sin \frac{k}{R}y
\ee
We also have
\be
\int_0^{2\pi R}dy~|D_\mu \phi|^2 &=&
\left(\p_\mu \eta^{(0)}\right)^2 + (e v)^2 \left(a_\mu^{(0)}+\frac{1}{ev}\p_\mu \sigma^{(0)}\right)^2 \nonumber\\
&+& \sum_{k=1}^\infty\left\{
\left(\p_\mu \eta^{(k)}\right)^2 + (e v)^2 \left(a_\mu^{(k)}+\frac{1}{ev}\p_\mu \sigma^{(k)}\right)^2\right\},\\
\int_0^{2\pi R}dy~|D_y \phi|^2 &=& - \sum_{k=1}^\infty\left\{\left(\frac{k}{R}\right)^2 \eta^{(k)2} + 
(ev)^2 \left(a_y^{(k)} -\frac{1}{ev} \frac{k}{R}\sigma^{(k)}\right)^2
\right\},
\ee
where we have used
\be
D_\mu \phi &=& \frac{1}{\sqrt{2\pi R}} \left\{\p_\mu \eta^{(0)} + i e v \left(a_\mu^{(0)}+\frac{1}{ev}\p_\mu \sigma^{(0)}\right)\right\} \nonumber\\
&+& \frac{1}{\sqrt{\pi R}} \sum_{k=1}^\infty
\left\{\p_\mu \eta^{(k)} + i e v \left(a_\mu^{(k)}+\frac{1}{ev}\p_\mu \sigma^{(k)}\right)\right\}\cos\frac{k}{R}y + \cdots,\\
D_y \phi &=& - \frac{1}{\sqrt{\pi R}} \sum_{k=1}^\infty \left\{\frac{k}{R} \eta^{(k)} - i ev \left(a_y^{(k)} 
- \frac{1}{ev} \frac{k}{R}\sigma^{(k)}\right)\right\}\sin\frac{k}{R}y + \cdots.
\ee
Finally, we have
\be
\int_0^{2\pi R} dy~ V = (\lambda v)^2 \eta^{(0)2} + (\lambda v)^2 \sum_{k=1}^\infty \eta^{(k)2}.
\ee
Putting everything  together, we find the quadratic Lagrangian
\be
{\cal L}_{4D}^{(2)} &=&
-\frac{1}{4} f^{(0)}_{\mu\nu}f^{(0)\mu\nu} + (e v)^2 \left(a_\mu^{(0)}+\frac{1}{ev}\p_\mu \sigma^{(0)}\right)^2 \nonumber\\
&+& \left(\p_\mu \eta^{(0)}\right)^2 - (\lambda v)^2 \eta^{(0)2}  
+ \sum_{k=1}^\infty
\left(\p_\mu \eta^{(k)}\right)^2 - \left[(\lambda v)^2 + \left(\frac{k}{R}\right)^2\right] \eta^{(k)2}
\nonumber\\
&-& \frac{1}{4} \sum_{k=1}^\infty f^{(k)}_{\mu\nu}f^{(k)\mu\nu} 
+ \frac{1}{2} \sum_{k=1}^\infty \left(\p_\mu a_y^{(k)} + \frac{k}{R} a_\mu^{(k)}\right)^2 \nonumber\\
&+& \sum_{k=1}^\infty
(e v)^2 \left(a_\mu^{(k)}+\frac{1}{ev}\p_\mu \sigma^{(k)}\right)^2 
- \sum_{k=1}^\infty 
(ev)^2 \left(a_y^{(k)} -\frac{1}{ev} \frac{k}{R}\sigma^{(k)}\right)^2.
\label{eq:L4d_AH}
\ee
One immediately sees that $\sigma^{(0)}$ and $a_\mu^{(0)}$ appear only in the combination  $a_\mu^{(0)}+\frac{1}{ev}\p_\mu \sigma^{(0)}$,
namely $\sigma^{(0)}$ is absorbed by $a_\mu^{(0)}$, so that $a_\mu^{(0)}$ gets non-zero mass by the ordinary Higgs mechanism.
In $R \to \infty$ limit, the KK tower becomes massless, and indeed all of $\sigma^{(k)}$ appears always with $a_\mu^{(k)}$. Namely,
infinite number of four-dimensional zero modes $\sigma^{(k)}$ are eaten by the infinite number of four-dimensional massless
gauge fields $a_\mu^{(k)}$. It is nothing but the Higgs mechanism in five dimensions.
In order to untangle the mixing at finite $R$, let us add the gauge fixing term of the Feynman gauge
\be
{\cal L}_{\rm F} &=& -\frac{1}{2}\int_0^{2\pi R} dy~\left(\p_M A^M - 2ev \sigma\right)^2 \nonumber\\
&=& -\frac{1}{2}\int_0^{2\pi R} dy~\left\{
\frac{1}{\sqrt{2\pi R}} \left(\p^\mu a^{(0)}_\mu - 2ev \sigma^{(0)}\right) + \frac{1}{\sqrt{\pi R}}\sum_{k=1}^\infty
\left(\p^\mu a^{(k)}_\mu - \frac{k}{R} a^{(k)}_y - 2ev \sigma^{(k)}\right)\cos\frac{k}{R}
\right\}^2 \nonumber\\
&=&- \frac{1}{2}\left(\p^\mu a^{(0)}_\mu - 2ev \sigma^{(0)}\right)^2 
- \frac{1}{2}\sum_{k=1}^\infty \left(\p^\mu a^{(k)}_\mu - \frac{k}{R} a^{(k)}_y - 2ev \sigma^{(k)}\right)^2\,,
\ee
where we have used
\be
\p^\mu A_\mu &=& \frac{1}{\sqrt{2\pi R}} \p^\mu a^{(0)}_\mu + \frac{1}{\sqrt{\pi R}} \sum_{k=1}^\infty \p^\mu a^{(k)}_\mu \cos\frac{k}{R}y\,,\\
\p^y A_y &=& - \frac{1}{\sqrt{\pi R}} \sum_{k=1}^\infty \frac{k}{R} a^{(k)}_y \cos \frac{k}{R}y\,.
\ee
In conclusion, we have
\be
{\cal L}_{4D}^{(2)} + {\cal L}_{\rm F} &=&
-\frac{1}{4} f^{(0)}_{\mu\nu}f^{(0)\mu\nu} - \frac{1}{2}\left(\p_\mu a^{(0)}_\mu\right)^2 + (e v)^2 \left(a_\mu^{(0)}\right)^2 \nonumber\\
&+& \left(\p_\mu \sigma^{(0)}\right)^2 - 2 (ev)^2 \sigma^{(0)2}
+ \sum_{k=1}^\infty\left\{ \left(\p_\mu\sigma^{(k)}\right)^2 - \left[2(ev)^2 + \left(\frac{k}{R}\right)^2\right]\sigma^{(k)2}\right\}
\nonumber\\
&+& \left(\p_\mu \eta^{(0)}\right)^2 - (\lambda v)^2 \eta^{(0)2}  
+ \sum_{k=1}^\infty
\left\{\left(\p_\mu \eta^{(k)}\right)^2 - \left[(\lambda v)^2 + \left(\frac{k}{R}\right)^2\right] \eta^{(k)2}\right\}
\nonumber\\
&+& \sum_{k=1}^\infty\left\{- \frac{1}{4} \sum_{k=1}^\infty f^{(k)}_{\mu\nu}f^{(k)\mu\nu} 
+ \frac{1}{2} \left[ 2(ev)^2 + \left(\frac{k}{R}\right)^2 \right]\left(a_\mu^{(k)}\right)^2 \right\}\nonumber\\
&+& \frac{1}{2}
\sum_{k=1}^\infty 
\left\{  \left(\p_\mu a_y^{(k)}\right)^2 -  \left[2(ev)^2 + \left(\frac{k}{R}\right)^2\right] \left(a_y^{(k)}\right)^2 \right\}\,.
\ee
Now we can read the mass spectrum as $\sqrt{2}\,ev$, $\sqrt{2(ev)^2 + (k/R)^2}$, $\lambda v$, and $\sqrt{(\lambda v)^2 + (k/R)^2}$.
Thus, all the masses are of order $M_5 = \{\lambda v, e v\}$ or higher than $M_5$. So, no lighter particles than $M_5$ do exist for any $R$.
This is, of course, because we compactify the fifth direction. Note that the masses shift to $0$,
 $k/R$, $\lambda v$, and $\sqrt{(\lambda v)^2 + (k/R)^2}$ when we turn off the gauge interaction $(e=0)$. The massless mode
 corresponds to NG for broken global $U(1)$ and the next lightest masses are $k/R$ at large $R$. In the scalar model (Goldstone model)
 those KK tower can be very light as $R$ increased, but once the gauge interaction turned on, their mass is lifted of order $M_5$.
 
 \renewcommand{\theequation}{B.\arabic{equation}}
\section{Identities for the effective Lagrangian}\label{app:A}

Calculation of the effective Lagrangians in 
Sec.~\ref{sc:effective_action2}
is greatly simplified by using a few useful identities described 
in this appendix. 

Let us first consider a generic integral appearing in the kinetic terms of scalar fields, namely
\begin{equation}
\lineint y\, \Tr\bigl[D_\mu f(y\mathbf{1}_5-\mathbf{Y})D^\mu f(y\mathbf{1}_5-\mathbf{Y})\bigr]  
\end{equation}
where ${\mathcal L}_{\mathbf{Y}} \equiv [\mathbf{Y},\cdot]$ is a Lie derivative.
The first step in evaluating this integral is to rewrite 
\begin{equation}
f(X) = f(\partial_\alpha)e^{\alpha X}\Big|_{a=0}\,,
\end{equation} 
which holds for any $f$ with a Taylor expansion around the origin. Since in all our calculations we only deal with entire functions, such as $\tanh(z)$ or $\sech(z)$, this is clearly satisfied. The second step involves the famous Poincar\'e identity
\begin{equation}
\delta e^{X}e^{-X} = \frac{e^{{\mathcal L}_X}-1}{{\mathcal L}_X}(\delta X)\,.
\end{equation}  
This leads to
\begin{gather}
\lineint y\, \Tr\bigl[D_\mu f(y\mathbf{1}_5-\mathbf{Y})D^\mu f(y\mathbf{1}_5-\mathbf{Y})\bigr] = \\
\lineint y\, \Tr\Bigl[f(\partial_\alpha)f(\partial_\beta)e^{(\alpha+\beta)(y\mathbf{1}_5-\mathbf{Y})}\frac{e^{\alpha{\mathcal L}_{\mathbf{Y}}}-1}{{\mathcal L}_{\mathbf{Y}}}\bigl(D_\mu \mathbf{Y}\bigr)\frac{1-e^{-\beta {\mathcal L}_{\mathbf{Y}}}}{{\mathcal L}_{\mathbf{Y}}}\bigl(D^{\mu}\mathbf{Y}\bigr)\Bigr]_{\alpha,\beta =0}
\end{gather}
Now we can formally shift the integration variable as $y\mathbf{1}_5-\mathbf{Y} \to y \mathbf{1}_5$. This can be established more rigorously by first diagonalising the matrix $e^{(\alpha+\beta)(y\mathbf{1}_5-\mathbf{Y})}$ and rewriting the integral as a sum of integrals for each diagonal element. We can then shift the integration variable to absorb each diagonal element of $\mathbf{Y}$ separately. Since there is no other $y$-dependent term in the above integral, this amounts to the shift  $y\mathbf{1}_5-\mathbf{Y} \to y \mathbf{1}_5$, as claimed.

Further, we will use the fact
\begin{equation}
f(\partial_\alpha) e^{\alpha y}\Big|_{\alpha = 0} = f(\partial_\alpha + y)\Big|_{\alpha = 0}
\end{equation}
and the properties of the trace to get
\begin{equation}
\lineint y\, \Tr\Bigl[f(\partial_\alpha+y\mathbf{1}_5)f(\partial_\beta+y\mathbf{1}_5)\frac{e^{\alpha{\mathcal L}_{\mathbf{Y}}}-1}{{\mathcal L}_{\mathbf{Y}}}\frac{e^{\beta{\mathcal L}_{\mathbf{Y}}}-1}{{\mathcal L}_{\mathbf{Y}}}\bigl(D_\mu \mathbf{Y}\bigr) D^{\mu}\mathbf{Y}\Bigr]_{\alpha,\beta=0}\,.
\end{equation}
This leads to the final result
\begin{equation}\label{eq:bosonidentity}
\lineint y\, \Tr\bigl[D_\mu f(y\mathbf{1}_5-\mathbf{Y})D^\mu f(y\mathbf{1}_5-\mathbf{Y})\bigr]  =\Tr\Bigl[g\bigl({\mathcal L}_{\mathbf{Y}}\bigr)\bigl(D_\mu \mathbf{Y}\bigr)D^{\mu}\mathbf{Y}\Bigr]\,,
\end{equation}
where
\begin{equation}
g(\alpha) \equiv \lineint y\,\Bigl(\frac{f(y+\alpha)-f(y)}{\alpha}\Bigr)^2\,.
\end{equation}

Let us also mention an identity relevant to our calculation of 
the moduli metric. If we decompose an adjoint field $Y$ as $Y = Y_i T_i$, where $T_i$ are generator of the $SU(N)$ algebra with the standard normalization
\begin{equation}
\comm{T_i}{T_j} = i f_{ijk}T_k\,, \hspace{5mm} \Tr\bigl[T_i T_j\bigr]=\frac{1}{2}\delta_{ij}\,,
\end{equation}
it is easy to show the following 
\begin{equation}\label{eq:gident}
g({\mathcal L}_{\mathbf{Y}})\bigl(T_i\bigr) = g\bigl(i \mathbf{\tilde Y}\bigr)_{ij}T_j\,,
\end{equation}
where $\mathbf{\tilde Y}_{ij} = f_{ijk}Y_k$. This simply comes out as a result of Taylor expanding the left side of the identity and repeatedly using the commutation relation for the generators.


\end{document}